\documentclass{aa}  
\usepackage{graphicx}
\usepackage{amsmath}
\usepackage{txfonts}
\usepackage{natbib}
\usepackage{url}

\begin{document}

\newcommand{\teff}{T$_{\rm eff}$}
\newcommand{\logg}{$\log${(g)}}
\newcommand{\met}{$[$Fe/H$]$}
\newcommand{\sind}{S$_{\rm \ion{Ca}{ii}}$}
\newcommand{\vsini}{$v\sin i$}
\newcommand{\hepsilon}{H$_{\rm\epsilon}$}
\newcommand{\hdelta}{H$_{\rm\delta}$}
\newcommand{\hgamma}{H$_{\rm\gamma}$}
\newcommand{\hbeta}{H$_{\rm\beta}$}
\newcommand{\halpha}{H$\rm\alpha$}

%\graphicspath{{images/}}
%
   \title{The HADES RV Programme with HARPS-N@TNG}

   \subtitle{IV. Time resolved analysis of the \ion{Ca}{II} H\&K and \halpha\ chromospheric emission of low-activity early-type M dwarfs}

   \author{G. Scandariato\inst{1}{\thanks{e-mail: gas@oact.inaf.it}}, J. Maldonado\inst{2}, L. Affer\inst{2}, K. Biazzo\inst{1}, G. Leto\inst{1}, B. Stelzer\inst{2}, R. Zanmar Sanchez\inst{1}, R. Claudi\inst{3}, R. Cosentino\inst{1,4}, M. Damasso\inst{5}, S. Desidera\inst{3}, E. Gonz\'alez \'Alvarez\inst{2,6}, J.I. Gonz\'alez Hern\'andez\inst{7,8}, R. Gratton\inst{3}, A.F. Lanza\inst{1}, A. Maggio\inst{2}, S. Messina\inst{1}, G. Micela\inst{2}, I. Pagano\inst{1}, M. Perger\inst{9}, G. Piotto\inst{10,4} R. Rebolo\inst{7,8}, I. Ribas\inst{9}, A. Rosich\inst{9,11,12}, A. Sozzetti\inst{4} and A. Su\'arez Mascare\~no\inst{7,8}}

   \institute{INAF -- Osservatorio Astrofisico di Catania, Via S.Sofia 78, I-95123, Catania, Italy
\and
INAF -- Osservatorio Astronomico di Palermo, Piazza del Parlamento, 1, I-90134, Palermo, Italy
\and
INAF -- Osservatorio Astronomico di Padova, Vicolo Osservatorio 5, 35122 Padova, Italy
\and
Fundaci\'on Galileo Galilei - INAF, Rambla Jos\'e Ana Fernandez P\'erez 7, 38712 Breaña Baja, TF, Spain
\and
INAF -- Osservatorio Astrofisico di Torino, Via Osservatorio 20, 10025, Pino Torinese, Italy
\and
Dipartimento di Fisica e Chimica, Universit\`a di Palermo, Piazza del Parlamento 1, I-90134 Palermo, Italy
\and
Instituto de Astrof\'isica de Canarias, 38205 La Laguna, Tenerife, Spain
\and
Universidad de La Laguna, Dpto. Astrofísica, 38206 La Laguna, Tenerife, Spain
\and
Instittut de Ci\`encies de l’Espai (IEEC-CSIC), Campus UAB, C/ Can Magrans, s/n, 08193 Bellaterra, Spain
\and
Dip. di Fisica e Astronomia Galileo Galilei – Universit\`a di Padova, Vicolo dell’Osservatorio 2, 35122, Padova, Italy
\and
Dept. d’Astronomia i Meteorologia, Institut de Ci\`encies del Cosmos (ICC), Universitat de Barcelona (IEEC-UB), Mart\'i Franqu\`es 1, 08028 Barcelona, Spain
\and
Reial Acad\`emia de Ci\`encies i Arts de Barcelona (RACAB), 08002 Barcelona, Spain
}
%   \date{Received September 15, 1996; accepted March 16, 1997}

% \abstract{}{}{}{}{} 
% 5 {} token are mandatory
 
  \abstract{M dwarfs are prime targets for current and future planet search programs, particularly of those focused on the detection and characterization of rocky planets in the habitable zone. In this context, understanding their magnetic activity is important for two main reasons: it affects our ability to detect small planets, and it plays a key role in the characterization of the stellar environment.}
  {We aim to analyze observations of the \ion{Ca}{ii} H\&K and \halpha\ lines as diagnostics of chromospheric activity for low-activity early-type M dwarfs.}
  {We analyze the time series of spectra of 71 early-type M dwarfs collected in the framework of the HADES project for planet search purposes. The HARPS-N spectra provide simultaneously the \ion{Ca}{ii} H\&K doublet and the \halpha\ line. We develop a reduction scheme able to correct the HARPS-N spectra for instrumental and atmospheric effects, and to provide also flux-calibrated spectra in units of flux at the stellar surface. The \ion{Ca}{ii} H\&K and \halpha\ fluxes are then compared with each other, and their time variability is analyzed.}
  {We find that the \ion{Ca}{ii} H and K flux excesses are strongly correlated with each other, while the \halpha\ flux excess is generally less correlated with the \ion{Ca}{ii} H\&K doublet. We also find that \halpha\ emission does not increase monotonically with the \ion{Ca}{ii} H\&K line flux, showing some absorption before being filled in by chromospheric emission when \ion{Ca}{ii} H\&K activity increases. Analyzing the time variability of the emission fluxes, we derive a tentative estimate of the rotation period (of the order of a few tens of days) for some of the program stars, and the typical lifetime of chromospheric active regions (of the order of a few stellar rotations).}
  {Our results are in good agreement with similar previous studies. In particular, we find evidence that the chromospheres of early-type M dwarfs could be characterized by different filaments coverage, affecting the formation mechanism of the \halpha\ line. We also show that chromospheric structure is likely related to spectral type.}


	\authorrunning{Scandariato et al.}

   \maketitle
%
%________________________________________________________________

\section{Introduction}

The term \lq\lq stellar activity\rq\rq\ indicates a class of phenomena that, in stars with an external convective envelope, are triggered by the reconfiguration of the surface magnetic field. For the Sun, these phenomena are commonly classified as, e.g.,  sunspots, plages, flares, coronal holes, and they take place from the photosphere to the corona. The portions of the stellar surface and atmosphere where these phenomena take place are referred to as Active Regions (ARs).

From the observational point of view, traditionally the most frequently investigated diagnostic of chromospheric activity for solar-type stars is the \ion{Ca}{ii} H\&K doublet \citep[3933.67 \AA\ and 3968.47 \AA, see e.g. the milestone project of the Mount Wilson Observatory as summarized in][]{Baliunas1998}. The main advantage of this doublet is that the photospheric line profiles have large depths with respect to the continuum, such that the chromospheric emission can be detected with high contrast against the photospheric background. Nonetheless, a particularly relevant drawback especially for M type stars is that detailed spectroscopy is observationally more demanding because of the low continuum level at wavelengths shortward of 4000 \AA. For this reason, these observations have been focusing on stars ranging from F to K spectral types.

This led to the identification of another diagnostic more suited for red dwarfs: the \halpha\ line \citep[6562.80 \AA, see e.g.][]{Pasquini1991,Montes1995}. It is generally accepted that the \ion{Ca}{ii} H\&K and \halpha\ excess fluxes are tightly correlated, despite the fact that they form in different layers of the chromosphere. However, this conclusion has been drawn historically by analyzing averaged measurements of chromospheric fluxes of calcium and hydrogen, taken at different epochs and rarely obtained simultaneously \citep{Giampapa1989,Robinson1990,Strassmeier1990}. Another approach is to collect simultaneous measurements for a number of stars at a given epoch, without any follow-up investigation of the observed targets at different epochs \citep{Thatcher1993,Walkowicz2009}.

Recently, planet search programs carried out with the radial velocity techniques started to monitor samples of M dwarfs. These stars are extremely interesting targets for planet discoveries. First of all, they represent $\sim$75\% of the stars in the solar neighborhood \citep{Reid2002,Henry2006}. Moreover, from an observational point of view, the chances of finding an Earth-like planet in the habitable zone of a star increase as the stellar mass decreases. Still, habitability is not guaranteed simply by an assessment of the distance from the star: several other factors, such as stellar activity, may move and/or shrink the habitability zone of a star \citep[see][and references therein]{Vidotto2013}. Thus, it is crucial to better understand the activity of M dwarfs and how it can affect the circumstellar environment.

The intensive monitoring aimed at an accurate measurement of radial velocities provided the community with large databases of high-resolution and high signal-to-noise ratio (S/N) optical spectra per star. These databases are valuable for the analysis of the activity of M dwarfs. For instance, \citet{Gomes2011} analyzed the long-term variability of a number of chromospheric optical indexes for a sample of 30 M0--M5.5 stars from the HARPS$@$ESO M-dwarf planet search program \citep{Bonfils2013} with a median timespan of observations of 5.2 years.

The HArps-N red Dwarf Exoplanet Survey \citep[HADES,][]{Affer2016,Perger2016} project is a collaborative program between the Global Architecture of Planetary Systems project\footnote{\url{http://www.oact.inaf.it/exoit/EXO-IT/Projects/Entries/2011/12/27_GAPS.html}} \citep[GAPS,][]{Covino2013}, the Institut de Ci\`encies de l'Espai (IEEC/CSIC), and the  Instituto de Astrof\'isica de Canarias (IAC), aiming at the monitoring of the radial velocities of a sample of low-activity M-type dwarfs to search for planets. In this framework, we are analyzing the collected spectra to study the activity of the monitored stars from different points of view. In \citet{Maldonado2016} we analyze the \ion{Ca}{ii} H\&K and Balmer line flux excesses in relation with rotation and age, while in \citet{rotation} we analyze the time series of spectra in order to measure the rotation period of the stars and the periodicity of their activity cycles, if any is present.

In this paper our aim is to characterize the daily-to-monthly variability due to chromospheric ARs based on the analysis of simultaneous measurements of the \ion{Ca}{ii} H\&K and \halpha\ lines. In Sect.~\ref{sec:data} we present the target selection and the database of spectra collected so far; in Sect.~\ref{sec:reduction} we describe the data reduction and the measurement of the chromospheric \ion{Ca}{ii} H\&K and \halpha\ flux excesses; in Sect.~\ref{sec:fluxflux} we analyze the flux-flux relationships and in Sect.~\ref{sec:variability} we discuss the amount and timescales of chromospheric variability in the selected sample of stars.

\section{Target selection}\label{sec:data}

The HADES sample of stars is made up of 78 red dwarfs, 71 of them observed so far. The observed stars have spectral types ranging between K7.5V and M3V (corresponding to the $\sim$3400--3900~K temperature range). These stars are being monitored with the HARPS-N spectrograph \citep{Cosentino2012} mounted at Telescopio Nazionale Galileo in the framework of the HADES project. The spectrograph covers the 383--693~nm wavelength range with a spectral resolution of $\sim$115,000. HARPS-N spectra were reduced using the most recent version of the Data Reduction Software (DRS) pipeline \citep{Lovis2007}.

The monitored stars have been selected as targets favorable for planet search, thus the sample is biased towards low activity levels with some exceptions. The program stars are listed in Table~\ref{tab:stars}, together with the number of observations for each of them. The data analyzed in this paper have been collected during 7 semesters, from September 2012 to February 2016. In \citet{Maldonado2016} we have measured the stellar parameters using the methodology described in \citet{Maldonado2015}. The most relevant stellar parameters for the current analysis are listed in Table~\ref{tab:stars}. We refer to \citet{Affer2016} for further details of the target selection and to \citet{Maldonado2016} for a more detailed description of the determination of the stellar parameters.

\longtab{
\begin{longtable}{lccccccc}
\caption{Relevant stellar parameters of the stellar sample, and number of HARPS-N spectra per target. See \citet{Maldonado2016} for a full description.}\label{tab:stars}\\
\hline
\hline
 Star   & T$_{\rm eff}$ &  Sp-Type   & [Fe/H]   &  $\log g$   &   $v\sin i$ & Age & N.obs.\\
        &   (K)         &            & (dex)    &  (cgs)	  & (km s$^{\rm -1}$) &  & \\ 
\hline
\endfirsthead
\caption{Continued.} \\
\hline
 Star   & T$_{\rm eff}$ &  Sp-Type   & [Fe/H]   &  $\log g$   &   $v\sin i$ & Age & N.obs.   \\
        &   (K)         &            & (dex)    &  (cgs)	  &  (km s$^{\rm -1}$) &  &  \\
\hline
\endhead
\hline
\endfoot
\hline
\endlastfoot
GJ~2				&	3713  &   M1		&   -0.14    &  4.76  & 	0.98   & Old & 95 \\
GJ~3014			&	3695  &   M1.5	&   -0.19    &  4.79  &   $<$ 1.08 & Old & 1\\
GJ~16				&	3673  &   M1.5	&   -0.16    &  4.78  & 	1.02   & Young & 107\\
GJ~15A			&	3607  &   M1		&   -0.34    &  4.87  & 	1.09   & Old & 88\\
GJ~21				&	3746  &   M1		&   -0.12    &  4.74  & 	1.46   & Young & 81\\
GJ~26				&	3484  &   M2.5	&   -0.17    &  4.88  &   $<$ 0.94 & Old & 39\\
GJ~47				&	3525  &   M2		&   -0.26    &  4.88  & 	1.81   & Young & 66\\
GJ~49				&	3712  &   M1.5	&   -0.03    &  4.73  & 	1.32   & Old & 95\\
GJ~1030			&	3658  &   M2		&   -0.08    &  4.76  &   $<$ 0.93 & Young & 4\\
NLTT~4188			&	3810  &   M0.5	&   -0.06    &  4.70  & 	1.11  & Young & 1 \\
GJ~70				&	3511  &   M2.5	&   -0.21    &  4.87  & 	1.02   &  Old & 20\\
GJ~3117A			&	3549  &   M2.5	&   -0.13    &  4.82  &   $<$ 0.91 & Old & 8\\
GJ~3126			&	3505  &   M3		&   0.01     &  4.80  &   $<$ 0.83 & Old & 2\\
GJ~3186			&	3768  &   M1		&   -0.14    &  4.74  &   $<$ 1.02 & Young & 1\\
GJ~119A			&	3761  &   M1		&   -0.08    &  4.72  & 	0.98   & Old & 89\\
GJ~119B			&	3508  &   M3		&   0.05     &  4.79  &   $<$ 0.81 & Old & 3 \\
TYC~1795-941-1	&	3774  &   M0		& 	0.01		 &  4.65  	& 	3.30  & Young & 1 \\
NLTT~10614		&	3728  &   M1.5	&   -0.06    &  4.73  & 	2.07   & Young & 1\\
TYC~3720-426-1	&	3822  &   M0		& 	0.12		 &  4.64	  	& 	4.13   & Young & 2\\
GJ~150.1B			&	3730  &   M1		&   -0.16    &  4.76  & 	0.87   & Old & 44\\
GJ~156.1A			&	3745  &   M1.5	&   -0.05    &  4.72  & 	2.85   & Old & 65\\
GJ~162			&	3746  &   M1		&   -0.19    &  4.77  & 	0.93   & Young & 64\\
GJ~1074			&	3765  &   M0.5	&   -0.16    &  4.75  & 	1.13   & Old & 26\\
GJ~184			&	3752  &   M0.5	&   -0.10    &  4.73  & 	1.45   & Old & 44\\
GJ~3352			&	3809  &   M0.5	&   -0.13    &  4.72  & 	1.47   & Old & 10\\
TYC~3379-1077-1	&	3896  &   M0		&   0.04     &  4.61  & 	1.85   & Old & 8\\
TYC~743-1836-1	&	3846  &   M0		&   -0.03    &  4.67  & 	1.73   & Young & 3\\
GJ~272			&	3747  &   M1		&   -0.19    &  4.77  & 	1.09   & Young & 9\\
StKM~1-650		&	3874  &   M0.5	&   -0.11    &  4.67  & 	1.12   & Old & 10\\
NLTT~21156		&	3616  &   M2		&   -0.05    &  4.77  & 	0.70   & Young & 36\\
GJ~399			&	3563  &   M2.5	&   0.15     &  4.72  &   $<$ 0.88 & Young & 30\\
GJ~408			&	3472  &   M2.5	&   -0.19    &  4.89  & 	0.97   & Young & 36\\
GJ~412A			&	3631  &   M0.5	&   -0.38    &  4.87  & 	1.20   & Old & 72\\
GJ~414B			&	3661  &   M2		&   -0.09    &  4.76  &   $<$ 0.94 & Old & 30\\
GJ~3649			&	3691  &   M1.5	&   -0.14    &  4.77  & 	1.55   & Young & 17\\
GJ~450			&	3649  &   M1.5	&   -0.20    &  4.80  & 	1.15   & Old & 22\\
GJ~9404			&	3875  &   M0.5	&   -0.10    &  4.67  & 	1.25   & Old & 32\\
GJ~476			&	3498  &   M3		&   -0.16    &  4.86  &   $<$ 0.93 & Old & 12\\
GJ~9440			&	3710  &   M1.5	&   -0.13    &  4.76  &   $<$ 0.99 & Young & 69\\
GJ~521A			&	3601  &   M1.5	&   -0.09    &  4.79  &   $<$ 0.90 & Old & 53\\
GJ~3822			&	3821  &   M0.5	&   -0.13    &  4.71  & 	0.98   & Young & 41\\
GJ~548A			&	3903  &   M0		&   -0.13    &  4.66  & 	1.11   & Old & 29\\
GJ~552			&	3589  &   M2		&   -0.09    &  4.79  &   $<$ 0.90 & Old & 61\\
GJ~606			&	3665  &   M1.5	&   -0.21    &  4.80  & 	1.57   & Old & 27\\
GJ~3942			&	3867  &   M0		&   -0.04    &  4.65  & 	1.67   & Young & 98\\
GJ~625			&	3499  &   M2		&   -0.38    &  4.94  & 	1.32   & Young & 99\\
GJ~3997			&	3754  &   M0		&   -0.24    &  4.78  & 	0.94   & Young & 63\\
GJ~3998			&	3722  &   M1		&   -0.16    &  4.77  & 	1.56   & Old & 140\\
GJ~2128			&	3518  &   M2.5	&   -0.30    &  4.90  & 	1.19   & Young & 16\\
GJ~671			&	3422  &   M2.5	&   -0.17    &  4.93  & 	0.91   & Old & 7\\
GJ~685			&	3816  &   M0.5	&   -0.15    &  4.72  & 	1.33   & Young & 44\\
GJ~686			&	3663  &   M1		&   -0.30    &  4.83  & 	1.01   & Old & 2\\
GJ~694.2			&	3847  &   M0.5	&   -0.21    &  4.72  & 	1.13   & Young & 105\\
GJ~4057			&	3873  &   M0		&   -0.15    &  4.69  & 	0.81   & Old & 86\\
GJ~720A			&	3837  &   M0.5	&   -0.14    &  4.71  & 	1.49   & Old & 78\\
GJ~731			&	3844  &   M0		&   -0.16    &  4.71  & 	1.59   & Old & 15\\
GJ~740			&	3845  &   M0.5	&   -0.14    &  4.70  & 	0.92   & Old & 80\\
GJ~4092			&	3858  &   M0.5	&   -0.06    &  4.67  & 	1.20   & Old & 13\\
GJ~9689			&	3824  &   M0.5	&   -0.13    &  4.71  & 	1.47   & Old & 58\\
GJ~793			&	3461  &   M3		&   -0.21    &  4.91  &   $<$ 1.00 & Old & 24\\
BPM~96441			&	3896  &   M0		&   -0.03    &  4.63  & 	2.05   & Young & 13\\
TYC~2710-691-1	&	3867  &   K7.5	&   0.02     &  4.63  & 	2.41   & Young & 1\\
TYC~2703-706-1	&	3822  &   M0.5	&   0.06     &  4.65  & 	3.32   & Young & 68\\
GJ~4196			&	3666  &   M1		&   0.07     &  4.71  & 	2.40   & Old & 1\\
NLTT~52021		&	3687  &   M2		&   -0.12    &  4.77  &   $<$ 0.97 & Old & 1\\
NLTT~53166		&	3832  &   M0		&   -0.11    &  4.70  & 	1.45   & Old & 11\\
GJ~9793 			&	3881  &   M0		& 	0.24		 &  4.58	  	& 	2.77   & Young & 30\\
GJ~4306			&	3763  &   M1		&   -0.13    &  4.74  & 	1.01   & Young & 123\\
GJ~895			&	3748  &   M1.5	&   -0.09    &  4.73  & 	1.70   & Old & 4\\
V*~BR~Psc			&	3553  &   M1.5	&   -0.29    &  4.88  & 	0.88   & Old & 10\\
%2MASSJ2235$^{\dag}$	& 3891 & K7.5	&	-0.13	 &  4.67  & 	2.04   & 4 Old &\\
2MASS~J22353504+3712131	& 3891 & K7.5	&	-0.13	 &  4.67  & 	1.92   & Old & 4\\
\end{longtable}
%\tablefoot{$^{\dag}$ 2MASS~J22353504+3712131}
}

%\fi

\section{Post-processing analysis}\label{sec:reduction}

During the spectra analysis, we find evidence of variability in the instrumental response, in the atmospheric reddening, and in the intensity of telluric lines, either in absorption or emission, on a night-to-night basis. These effects are usually corrected by observing a spectroscopic standard star and a telluric standard star close to the observed target. The aim of the HADES and GAPS projects is to find new exoplanets around pre-selected targets or to refine the characterization of known exoplanetary systems. Thus, the instrumental setup and the observational campaign are arranged to maximize the number of targeted stars and the precision of the radial velocity measurements, while the flux calibration of the spectra is not taken into account (e.g.\ neither spectroscopic or telluric standard stars are usually observed).

For our purposes, however, the correction of all these effects is necessary in order to compare the \ion{Ca}{ii} H\&K and \halpha\ lines taken in different nights. In this section we describe in detail our post-processing analysis of the spectra reduced with the DRS pipeline.

\subsection{Flux rescaling}\label{sec:rescale}

%gsGenerateRdbs.pro
%input: HARPS spectra
%output: file.path('output',stars[s],'asymmetry_analysis_bi_gaussian.dat')

%gsCorrectSpectra.r
%input: file.path('output',stars[s],'asymmetry_analysis_bi_gaussian.dat')
%output: file.path('output',stars[s],'corrected.sav')

\begin{figure}
\centering
\includegraphics[viewport=10 8 280 432,clip,width=.9\linewidth]{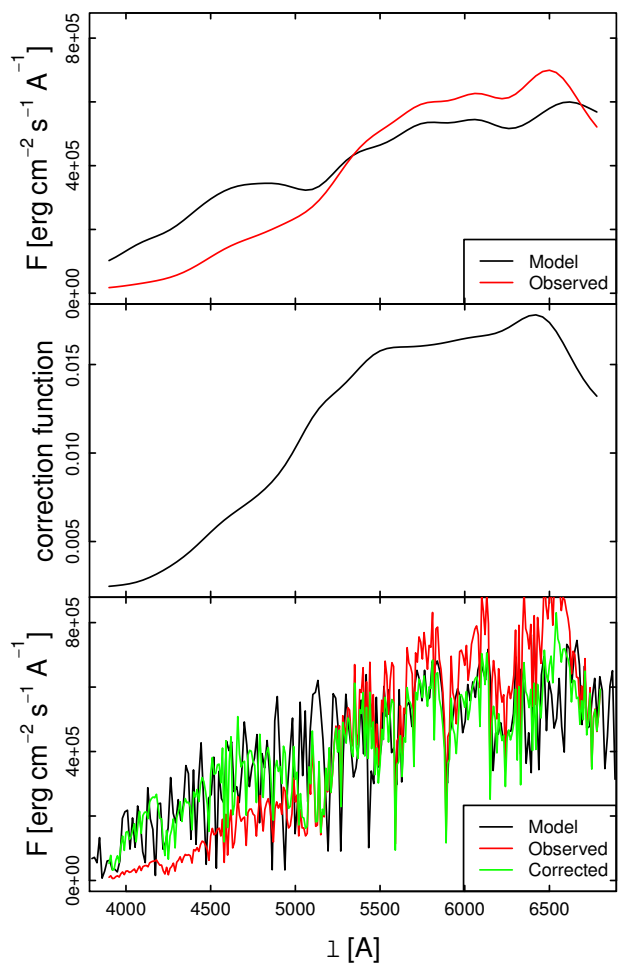}\\
\caption{Example of the correction of the observed spectra discussed in the text. \textit{Top panel - }Comparison between the BT-Settl model spectrum (in black) and the observed spectrum (in red), both degraded to low resolution. For plotting reasons, the observed spectrum, still in instrumental units, is rescaled to approximately match the flux scale of the model spectrum. \textit{Middle panel - }Observed-to-model ratio correction. The curve describes the rescaling of the observed spectrum into absolute units. \textit{Bottom panel - }Result of the rescaling: the high resolution spectra corresponding to the ones shown in the top panel are plotted in black and red again. The green line shows the corrected high resolution spectrum.}\label{fig:dereddening}
\end{figure}

To correct the observed fluxes for instrumental response and atmospheric reddening, for each star in our catalog we interpolate the synthetic spectral library provided by \citet{Allard2011}\footnote{We adopt the CIFIST2011 models (\url{https://phoenix.ens-lyon.fr/Grids/BT-Settl/CIFIST2011bc/SPECTRA/})%, which use the \citet{Caffau2011} solar abundances and take into account the calibration of the mixing length based on RHD simulations by \citet{Freytag2010} with additional adjustments to the MLT equations.
} to compute a model spectrum corresponding to the measured \teff, \logg\ and \met. We arbitrarily degrade the spectral resolution of both the model spectrum and the series of observed spectra convolving them with a gaussian kernel with $\sigma$=120~\AA, corresponding to a final spectral resolution of $\sim$50 over the spectral range in consideration (Fig.~\ref{fig:dereddening}, top panel). The ratio of each low-resolution spectrum and the model is thus a continuous function of wavelength (Fig.~\ref{fig:dereddening}, middle panel), which allows us to rescale the high-resolution instrumental fluxes to fluxes at stellar surface and, at the same time, to correct for the differential instrumental response and atmospheric reddening (Fig.~\ref{fig:dereddening}, bottom panel).

The degradation to such a low spectral resolution is motivated by different reasons:
\begin{itemize}
\item it allows us to safely mask out the narrow spectral windows centered on the chromospheric emission lines (Table~\ref{tab:indicators}), which may introduce spurious features in the correction function; 
\item it removes the line component of the telluric spectrum, leaving us only with the continuous component of the atmospheric absorption;
\item at low resolution, the discrepancies between observed and model line profiles are negligible, the low resolution spectra being dominated by the continuum flux.
\end{itemize}

%gsCorrectTelluric
%input: file.path('output',stars[s],'corrected.sav')
%output: file.path('output',stars[s],'cat.sav')

\subsection{Telluric line removal}\label{sec:telluric}

The other effect to take into account is the telluric contamination. We focus only on the 130~\AA-wide spectral range around the \halpha\ line, because the \ion{Ca}{ii} H\&K spectral region is free from telluric lines.

In order to have a comparison spectrum suited for the removal of telluric lines, for each star we compute the median of all the flux-rescaled high-resolution spectra. This median spectrum, besides having higher S/N compared to the single spectra, is also free from telluric lines. As a matter of fact, telluric lines are Doppler shifted in the reference frame of the targeted star for the observations are generally performed at different hour angles and different months along the observing seasons. Since their wavelengths change in the stellar rest frame from night to night, the median algorithm filters them out as outliers.

For each star we compute the ratio between each spectrum and the corresponding median one, obtaining the normalized telluric spectrum affecting each given observation. Then we cross-correlate it with the normalized high-S/N spectrum of $\eta$ UMa, a telluric standard star we observed with HARPS-N within the GAPS program \citep[see][for further details]{Borsa2015}. While the cross-correlation allows to align this model telluric spectrum to the ratio of the spectra, the intensity of the telluric contamination is computed by a 3-$\sigma$ clipped comparison between the wavelength-shifted model telluric spectrum and the ratio. These two parameters (wavelength shift and intensity) allow us to correct the telluric lines within the noise of the observed spectra.

In the following sections, we will always mean the flux-rescaled telluric-corrected spectra when referring to the observed spectra.

\subsection{Selection of the reference spectra}\label{sec:referenceSpectra}

%gsChromosphericExcessNew.r
%input: file.path('output',stars[s],'cat.sav')
%output: file.path('output',stars[s],'cat.sav')

%file.path('savs','indexCollection.sav') is the output of gsPlotIndeces.r

For each star, we have series of individual spectra and the corresponding median spectrum (Sect.~\ref{sec:rescale} and Sect.~\ref{sec:telluric}). To make the computation faster, we extract a 70~\AA\ spectral interval around the \halpha\ and a 100~\AA\ interval around the \ion{Ca}{ii} H\&K doublet comprising both lines (Table~\ref{tab:indicators}).

In each interval, we want to subtract the photospheric flux to measure the chromospheric flux. To this purpose, our first approach was to use synthetic spectra, which soon turned out to loosely fit the absorption line profiles. We then decided to use our spectral database to obtain the best \lq\lq inactive\rq\rq\ template for each star in the program.

\begin{figure}
\centering
\includegraphics[width=.9\linewidth,viewport=1 10 350 260,clip]{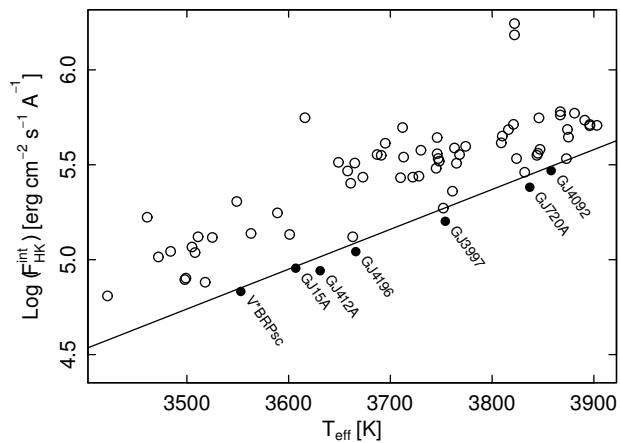}
\caption{$F_{HK}^{int}$ vs. \teff\ diagram. The solid line (Eq.~\ref{eq:selection}) represent our selection threshold of reference stars. The selected stars are indicated as plain dots and their names are shown.}\label{fig:refSpectra}
\end{figure}

%gsMeasureLineFluxNoSubtraction.r
It is reasonable to assume that those stars with minimum \ion{Ca}{ii} H\&K emission are the \lq\lq least active\rq\rq\ and therefore the ones that should be used as templates for the spectral subtraction. For each star we measure the \ion{Ca}{ii} H\&K flux $F^{int}_{HK}$ integrating the median spectrum in the spectral intervals around the \ion{Ca}{ii} H and K lines defined in Table~\ref{tab:indicators} (see the end of Sect.~\ref{sec:spectralAnalysis} for the definition of the spectral intervals). These integrated fluxes thus contain the contribution of both the photospheric flux and chromospheric basal flux, plus the flux from chromospheric ARs if present. In Fig.~\ref{fig:refSpectra} we plot the integrated \ion{Ca}{ii} H\&K flux versus the stellar effective temperature: we find that the lower boundary of the diagram is an increasing function of \teff. This trend is due to the fact that the \ion{Ca}{ii} H\&K photospheric flux of early-type M dwarfs increases with \teff, as predicted by the BT-Settl models.

We select the template stars at the lower boundary (i.e.\ at minimum \ion{Ca}{ii} H\&K flux) of the diagram in Fig.~\ref{fig:refSpectra}, arbitrarily considering the stars below the line with equation
\begin{equation}
\rm Log(F_{HK}^{int})=-2.61+0.0021\cdot T_{eff}.\label{eq:selection}
\end{equation}
The selected reference stars are V$^*$~BR~Psc, GJ~15A, GJ~412A, GJ~4196, GJ~3997, GJ720A and GJ~4092. All these stars have sub-solar metallicities, with the exception of GJ~4196. The selection of these stars could reflect a trend between \ion{Ca}{ii} flux and metallicity due to the dependence of the \ion{Ca}{ii} line core strength on the calcium elemental abundance. This correlation could therefore be linked to the line core strength curve of growth in the atmosphere discussed by \citet{Houdebine2011}.

As we will discuss below, stellar metallicity cannot be neglected in the spectral subtraction. For this reason, we reject GJ~4196 as a reference star, and we keep the remaining stars as a spectral grid for the computation of the reference quiet spectrum.

\subsection{Spectral subtraction and line flux measurement}\label{sec:spectralAnalysis}

For each program star, the reference spectrum is computed interpolating the grid of reference spectra over \teff. Before the interpolation takes place, we broaden the spectra such that they all have the same rotational broadening. For this purpose, we select the maximum \vsini\ among those of the program and reference star as measured by \citet{Maldonado2016}, and we broaden each spectrum to this maximum \vsini\ using the rotational profile provided by \citet{Gray1992}.

\begin{figure}
\centering
\includegraphics[width=\linewidth,viewport=60 1150 615 1423,clip]{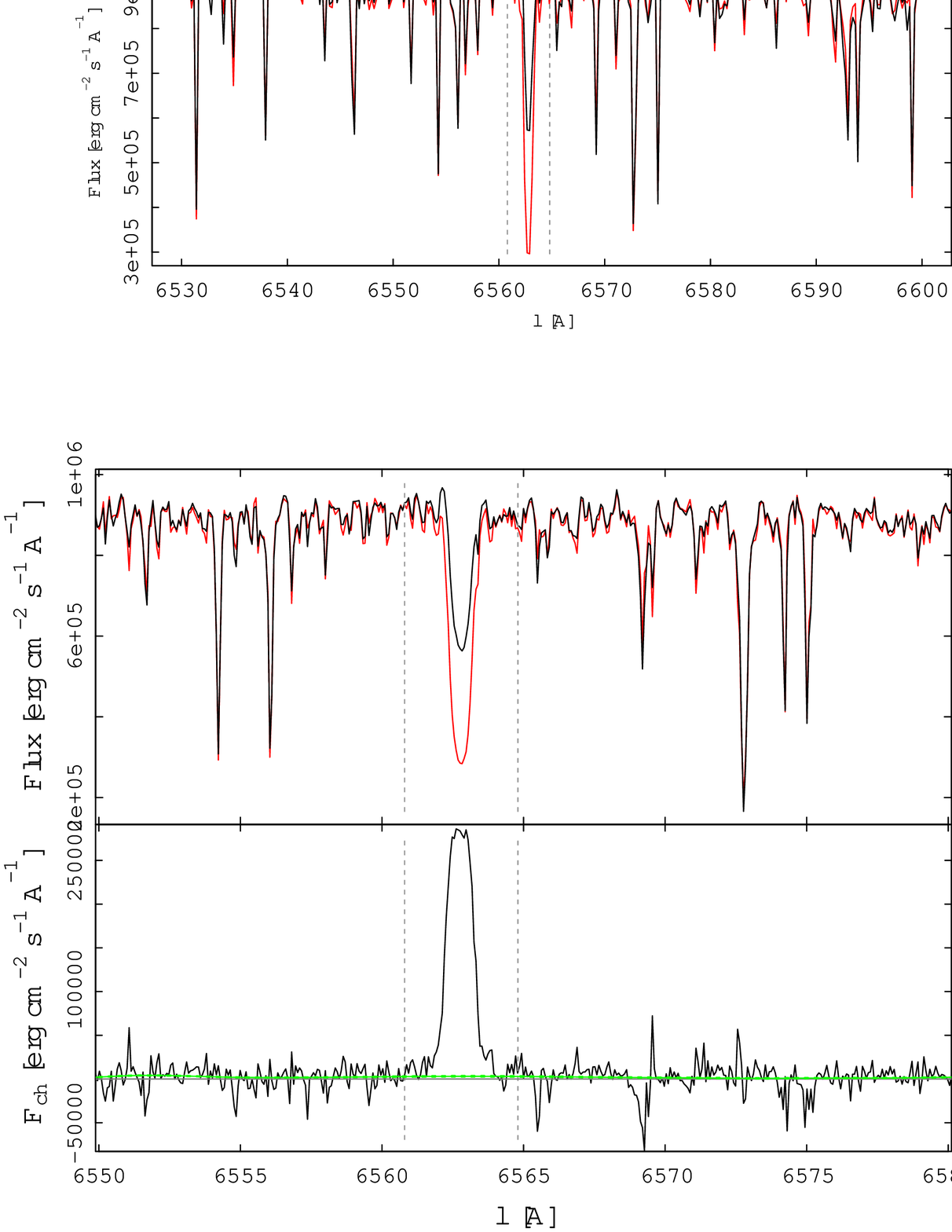}
\caption{Comparison between fluxes in the \halpha\ region of the spectrum of GJ~9793 and the corresponding interpolated reference spectrum (see text). Gray crosses represent fluxes in the \halpha\ core, and are not taken into consideration in the flux rescale. The red solid line is the 1:1 line. The green solid line is the fitted function used for the flux rescaling.}\label{fig:rescaleMet}
\end{figure}

\begin{figure}
\centering
\includegraphics[width=\linewidth,viewport=25 40 630 525,clip]{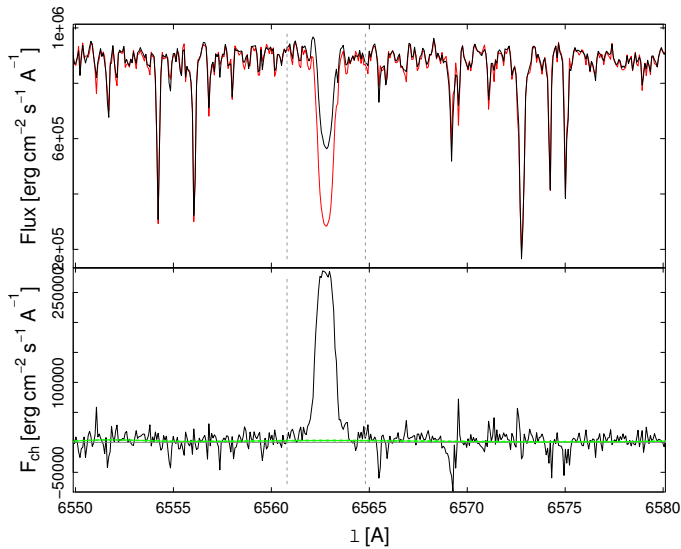}
\caption{Example of the extraction of the \halpha\ line flux. \textit{Top panel - }Zoom-in of the spectral comparison between the observed spectrum (in black) and the rescaled template (in red). The vertical gray dashed lines bracket the line core. \textit{Bottom panel - }Chromospheric spectrum, computed as the difference between the observed and template spectrum shown in the top panel. The green line is the polynomial fit for the removal of low-order trends in the differential spectrum.}\label{fig:example}
\end{figure}

Since the reference spectrum is computed using the spectra of stars with minimum metallicities in our sample, we find a systematic offset (see Fig.~\ref{fig:rescaleMet}) correlated with the metallicity of the star. In particular, this is due to the fact that the continuum flux predicted by the BT-Settl models decreases with increasing \met, consistently with the increase of the absorption by the multitude of atomic and molecular lines in M star spectra. In Fig.~\ref{fig:rescaleMet} we show the example of the \halpha\ spectrum of GJ~9793 (see also Fig.~\ref{fig:example}), the highest metallicity star in our sample: its spectrum shows lower fluxes compared with the corresponding reference spectrum. To correct this offset, we fit a low order polynomial curve to the flux points. In the fit, we exclude the core of the line (gray crosses in the figure), because it may be affected by chromospheric emission. Finally the reference spectrum is rescaled to the observed spectrum using the best-fit curve.

Once the reference spectrum is rescaled to the observed spectrum, the chromospheric spectrum is computed by subtraction of the template \citep[see also][for previous application of the spectral subtraction technique]{Herbig1985, Frasca1994}. Outside of the line cores we sometimes find a residual wavelength-dependent offset significantly different from zero. This residual offset arises from several factors, such as inaccuracies in the measured stellar parameters, the interpolation of the reference spectrum, and spectral peculiarities of the analyzed spectrum. To remove these systematics, we subtract a low-order polynomial function from the chromospheric spectrum, fitted over the spectral range excluding the line cores. We remark that the 1$\sigma$ confidence band on the polynomial fit is narrow enough to introduce a negligible uncertainty on the line flux compared to the noise (Fig.~\ref{fig:example}, bottom panel).

The chromospheric emission is finally computed integrating the chromospheric spectrum, i.e. the template-subtracted systematics-corrected spectrum, over the wavelength ranges listed in Table~\ref{tab:indicators}. The widths of the spectral windows used to compute the emission flux are conveniently set after visual inspection of the differential spectra, representing a compromise between the bracketing of the whole flux radiated by the chromosphere and the exclusion of the nearby spectral regions, which introduce noise in the flux measurement. To compute the uncertainty on the integrated flux, we measure the S/N of the differential spectrum outside of the line cores, and we propagate it in the integration of the line flux. This error thus includes both the noise in the observed spectra and the uncertainties introduced by our data reduction.

We caution that the spectral subtraction removes the flux of both the photosphere and the \lq\lq quiet\rq\rq\ chromosphere from the observed spectra. For the \ion{Ca}{ii} H\&K doublet, this means that the measured emission fluxes do not include the basal chromospheric emission, i.e.\ they are representative of the excess chromospheric flux radiated by ARs. The case of the \halpha\ line is more subtle. Following the chromospheric models of \citet{Cram1987}, the absorption in the \halpha\ line seen in the spectra of early-type M dwarfs is originated in the chromosphere. In particular, in these models even the unperturbed chromospheres are hot enough to populate the N=2 levels of the hydrogen ions, making the absorption of the \halpha\ line possible. Thus, in the least active M stars in our sample we see a \lq\lq basal\rq\rq\ \halpha\ absoption depth, and the line fluxes measured with the spectral subtraction technique have this basal absorption as zero point.

\begin{table}
 \begin{center}
\caption{Chromospheric emission lines analyzed in this work.}\label{tab:indicators}
  \begin{tabular}{lccc}
 \hline\hline
Ion & Line & Central wavelength (\AA) & Spectral width (\AA)\\
\hline
\ion{Ca}{ii} & K & 3933.67 & 1.5\\
\ion{Ca}{ii} & H & 3968.47 & 1.5\\
\hline
%\ion{H}{i} & \hepsilon & 3970.07 & 1.5\\
%\ion{H}{i} & \hdelta & 4101.76 & 1.5\\
%\ion{H}{i} & \hgamma & 4340.46 & 1.5\\
%\ion{H}{i} & \hbeta & 4861.32 & 3\\
\ion{H}{i} & \halpha & 6562.80 & 4\\
\hline
  \end{tabular}
 \end{center}
\end{table}

%%%%%%%%%%%%%%%%%%%%%%%%%%%%%%%%%%%
%\input{periods}
%%%%%%%%%%%%%%%%%%%%%%%%%%%%%%%%%%%

\section{Flux-flux relationships}\label{sec:fluxflux}
%%gsPlotIndeces.r, then gsAnalyzeVariabilitySeasonalMedian.r
%input: file.path('output',stars[s],'cat.sav')

In the following we analyze the relationships between the chromospheric fluxes in the \ion{Ca}{ii} H\&K and \halpha\ lines resolved in time. To improve the statistical reliability of our analysis, we arbitrarily discard all the stars with fewer than 20 observations, which leaves us with 41 stars. Moreover, to clean up our samples of measurements, for each star and each line we reject the outliers with a 5$\sigma$-clipping criterion. We apply the same selection to the measurement uncertainties, thus discarding the data with anomalously large error bars.% 

\subsection{\ion{Ca}{ii} H vs. \ion{Ca}{ii} K}\label{sec:caHK}

\begin{figure}
\centering
\includegraphics[width=.9\linewidth,viewport=4 8 350 265,clip]{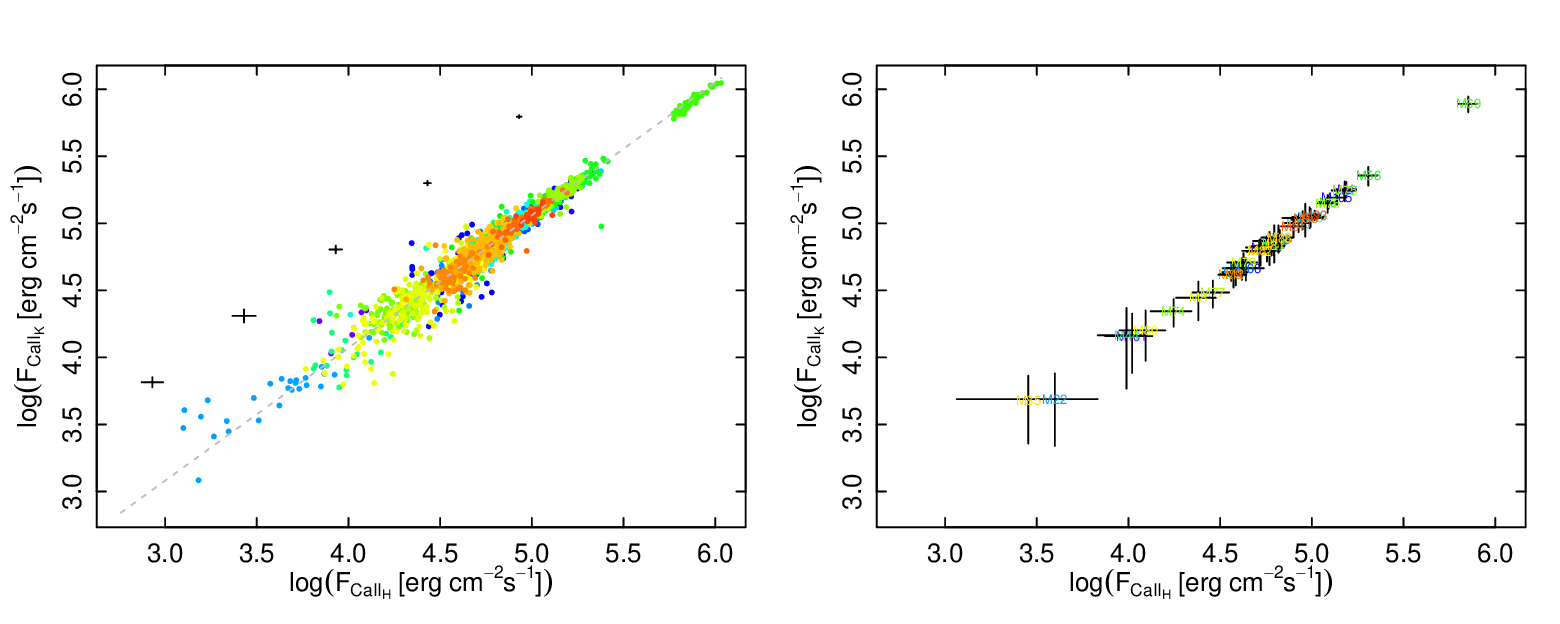}
\caption{$\rm\log(F{_{CaII_H}})$ vs. $\rm\log(F{_{CaII_K}})$. Different colors represent different stars. The gray dashes show the RMA best fit in Eq~\ref{eq:coeffHK}. Black crosses represent the typical measurement uncertainties at different activity levels.}\label{fig:HK}
\end{figure}

In Fig.~\ref{fig:HK} we compare the individual measurements of the \ion{Ca}{II} H and K excess fluxes (F$\rm{_{CaII_H}}$ and F$\rm{_{CaII_K}}$ respectively) of the stars in the analyzed sample, omitting the measurements less than 3$\sigma$ above zero, which are not representative of flux in excess. We find that the fluxes are well aligned despite differences in stellar parameters and activity levels. Since the uncertainties on the quantities are comparable, we compute the best-fit line by means of Ranged Major Axis (RMA) regression \citep{Legendre1998}. The result of the fit is
\begin{equation}
\log F_{\ion{Ca}{ii}_K}=(0.11\pm0.04)+(0.990\pm0.009)\log F_{\ion{Ca}{ii}_H},\label{eq:coeffHK}
\end{equation}
which is consistent within 3$\sigma$ with similar previous studies performed on samples of M dwarfs \citep{Martinez2011,Stelzer2012}.

From Eq.~\ref{eq:coeffHK} we derive that the K/H flux ratio is close to 1, thus following \citet{Houdebine1997} we argue that the emission of the \ion{Ca}{II} H\&K lines is optically thick.

\begin{figure}
\centering
\includegraphics[width=.9\linewidth,viewport=21 6 350 325,clip]{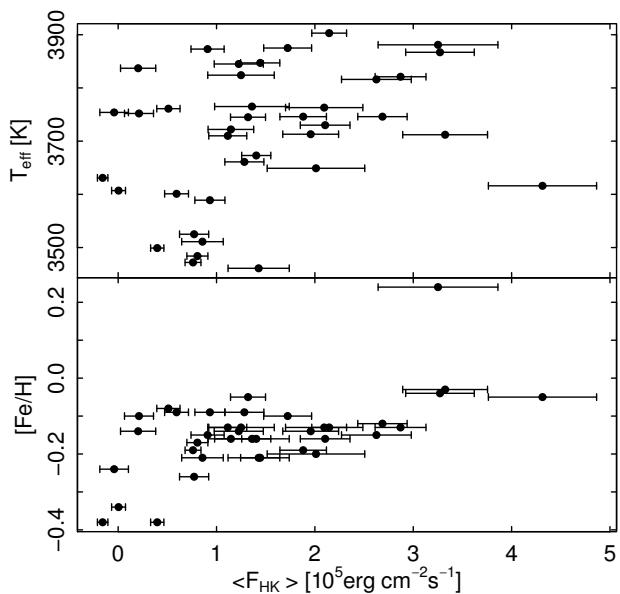}
\caption{\teff\ and \met\ versus the median total flux excess $\rm<F_{HK}>$. Error bars show the Median Absolute Deviation of the sample of $\rm F_{HK}$ measurements for each star.}\label{fig:actTeffMet}
\end{figure}

We analyze the \ion{Ca}{ii} H\&K flux excess as a function of temperature and metallicity. For this purpose, in Fig.~\ref{fig:actTeffMet} we plot the median total flux excess $\rm<F_{HK}>=<F_H+F_K>$ of the stars, together with the corresponding Median Absolute Deviation\footnote{The Mean Absolute Deviation (MAD) is a measure of the scatter of a sample of data. It is defined as the median of the absolute residuals with respect to the sample median, and is thus robust against the presence of outliers.} (MAD) of the sample of measurement, versus \teff\ and \met.

We find indication that the flux excess drops for stars cooler than \teff=3600~K (spectral type M1). Similar results have been found by \citet{West2004} analyzing the \halpha\ emission, and by \citet{West2011} including the \ion{Ca}{ii} K line, but for stars later than M4. We cannot claim any stronger conclusion because we are investigating a narrow spectral range. Moreover, we remark that our sub-sample of 41 stars is biased towards low activity levels, thus our findings are strongly affected by selection effects.

The dependence on the stellar metallicity is slightly more convincing, more metallic stars tending towards larger $\rm<F_{HK}>$ excess. One possible explanation is that the \ion{Ca}{ii} H\&K emission increases with the abundance of \ion{Ca}{ii} ions in the chromosphere, thus with metallicity at first approximation, as also discussed by \citet{Houdebine2011}. Another possibility is that this could be an age effect, as higher metallicities generally corresponds to younger ages. This is consistent with the fact that younger stars tend to be more active, as we discuss in Sect.~\ref{sec:varactage}.

\subsection{\ion{Ca}{ii} H\&K vs. \halpha}\label{sec:HKHalpha}

\begin{figure*}
\centering
\includegraphics[viewport=13 50 350 280,clip,width=.32\linewidth]{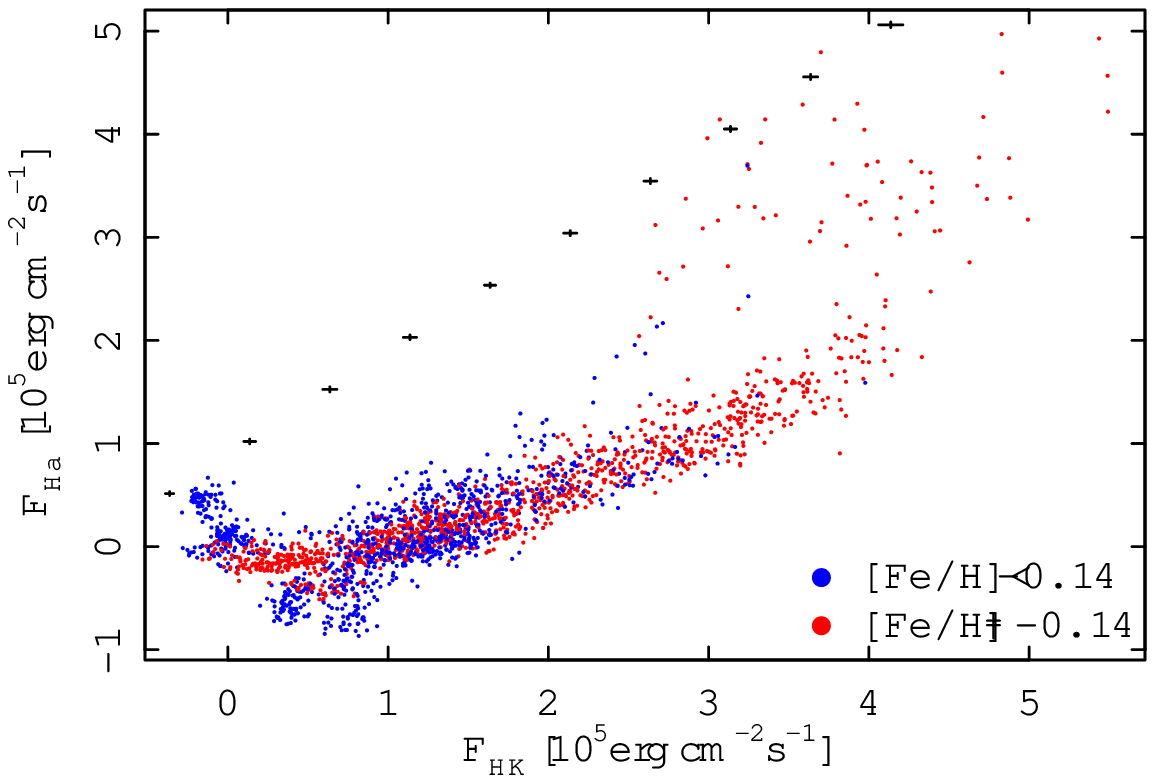}
\includegraphics[viewport=13 50 350 280,clip,width=.32\linewidth]{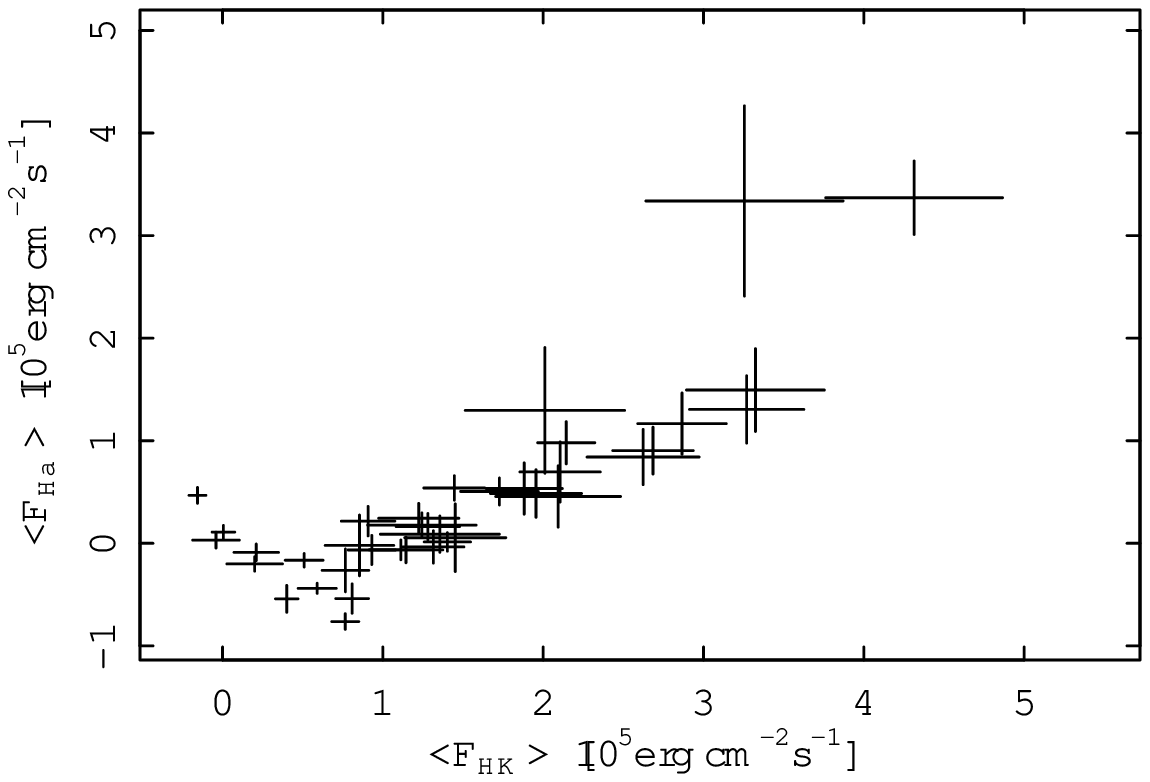}
\includegraphics[viewport=13 50 350 280,clip,width=.32\linewidth]{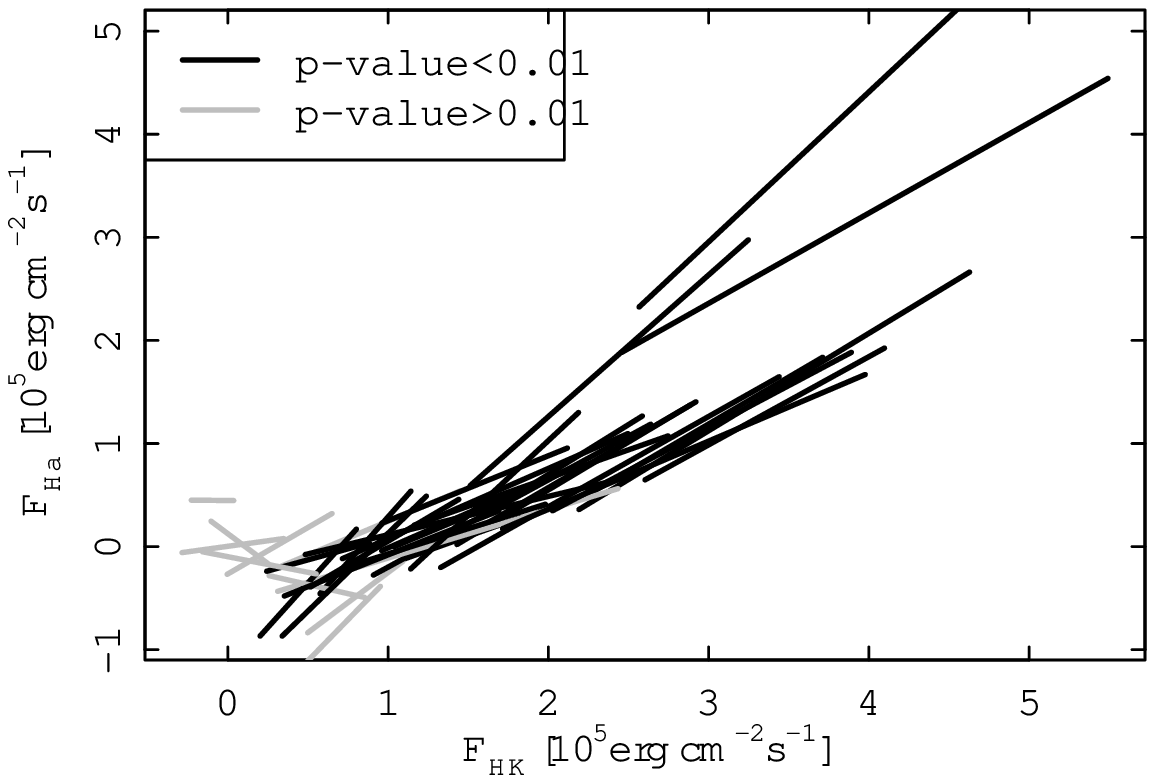}
\caption{\textit{Left panel - }Close-up view of the $\rm F_{HK}$ vs. $\rm F_{H\alpha}$ diagram. Colors code the metallicity of the stars as shown in the legend. Black crosses represent the typical measurement uncertainties at different activity levels. \textit{Middle panel - }Same as in the left panel, for the median of the flux measurements. For each star, the error bars represent the MAD of the measurements shown in the left panel. \textit{Right panel - }The same as in the left panel, representing each star with the corresponding linear fit to its flux-flux values. Gray lines mark best fits with low statistical significance, while black lines represent p-values lower than 1\% based on Spearman's correlation test.
}\label{fig:HalphaHK}
\end{figure*}

In the left panel of Fig.~\ref{fig:HalphaHK} we show the emission flux $\rm F_{H\alpha}$ radiated in the \halpha\ line against the flux $\rm F_{HK}$ radiated by the \ion{Ca}{ii} H\&K doublet. The comparison between \halpha\ and \ion{Ca}{ii} H\&K fluxes is more scattered than that of \ion{Ca}{ii} H and K lines. To avoid cluttering, in the middle panel we show the same diagram for the median of the flux measurements $\rm<F_{HK}>$ and $\rm<F_{H\alpha}>$, where the plotted uncertainties represent the MAD of the individual flux measurements.

One possible explanation of the scatter in Fig.~\ref{fig:HalphaHK} is that, having a larger number of \ion{Ca}{ii} ions in the atmosphere, stars with higher metallicities have more efficient \ion{Ca}{ii} H\&K radiative cooling (i.e.\ higher fluxes in emission), as found by \citet{Houdebine2011}. To test this hypothesis, we split the sample of stars into two halves at \met=-0.14, and plot them with different colors in the left panel of Fig.~\ref{fig:HalphaHK}, finding that the two subsamples follow the same locus. In particular, we do not find that, at fixed $\rm F_{H\alpha}$, the $\rm F_{HK}$ fluxes tend to increase with metallicity. We thus conclude that we do not have enough statistical evidence to state that metallicity affects the \halpha\ vs. \ion{Ca}{ii} H\&K excess relationship, maybe because the metallicity range spanned by our program stars is not large enough (see Table~\ref{tab:stars}).

\begin{figure}
\centering
\includegraphics[viewport=13 50 350 280,clip,width=.8\linewidth]{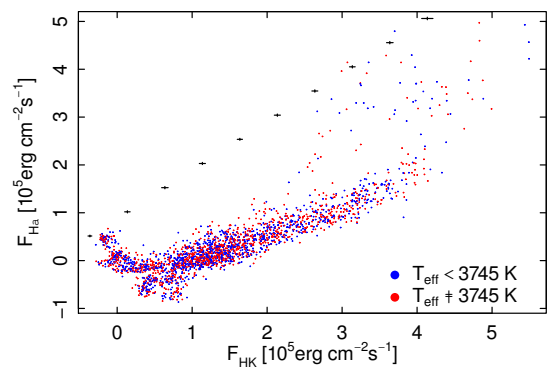}
\caption{Same as in the left panel of \ref{fig:HalphaHK}, splitting the analyzed sample in two halves at \teff=3745~K.}\label{fig:HalphaHKteff}
\end{figure}

We perform the same analysis splitting the sample of stars in two halves on the basis of their \teff\ (Fig.~\ref{fig:HalphaHKteff}). Also in this case, we do not find any evidence of correlation, maybe due to the fact that we are investigating a narrow range in \teff.

Another interesting feature, visible in the middle panel of Fig.~\ref{fig:HalphaHK}, is that the flux-flux relationship is not monotonic. As a matter of fact, a visual analysis of the plot shows that the flux radiated in the \halpha\ line seems to initially decrease with increasing \ion{Ca}{ii} H\&K excess flux, then goes below zero at $\rm<F_{HK}>\simeq10^{-5}erg\ cm^{-2} s^{-1}$ (i.e.\ the \halpha\ line  deepens), and finally increases with the \ion{Ca}{ii} H\&K flux. Similar results were obtained by \citet{Robinson1990}. The absorption of the \halpha\ line towards negative $\rm<F_{H\alpha}>$ values is the reason why we plot the \ion{Ca}{ii} H\&K vs.\ \halpha\ relationship in linear rather than logarithmic scale.

The non-monotonic trend is supported by statistical tests. Kendall's correlation test\footnote{Kendall's rank coefficient is often used as a non-parametric test statistic to establish whether two variables are statistically dependent \citep{Kendall1938}.} significantly supports that the two datasets are correlated. Moreover, we naively fit a linear and a quadratic function to the data, and we find that the corrected Akaike Information Criterion\footnote{Given a collection of models for the data, the Akaike Information Criterion \citep{Burnham2002} estimates the quality the models relative to each other. Hence, it provides a means for model selection.} significantly rejects the former compared to the latter.

From the theoretical point of view, this result has been predicted by chromospheric models \citep{Cram1979,Cram1987,Rutten1989,Houdebine1995,Houdebine1997}. According to these models, the most feasible scenario is that the chromospheric \ion{Ca}{ii} H\&K emission lines are collisionally-dominated, thus the radiated flux steadily increases with pressure. Conversely, the \halpha\ line is radiation-dominated and the increase of the optical depth initially leads to a deeper absorption profile, until the electron density is high enough to take the \halpha\ line into the collisionally-dominated formation regime, leading to the fill-in of the line.

Previous observational studies performed on similar samples of stars \citep[K and M type main sequence stars, e.g.][]{Stauffer1986,Giampapa1989,Rauscher2006,Walkowicz2009} have found comparable results. In particular, while the deepening of the \halpha\ line is still somehow controversial, it is clear that the \halpha\ is not filled in until a certain \ion{Ca}{ii} H\&K emission flux is reached. Thus, the \halpha\ line is not a good activity indicator for low to intermediate activity levels. Moreover, \citet{Houdebine1997} remark that observational data do not strictly match the models likely due to surface inhomogeneities, which may also explain the scatter in Fig.~\ref{fig:HalphaHK}.

A careful analysis of the left panel of Fig.~\ref{fig:HalphaHK} also shows that the stars follow individual loci with different slopes in the diagram. To better show this effect, in the right panel of Fig.~\ref{fig:HalphaHK} we plot the locus occupied by each single star defined as the RMA best fit (see Sect.~\ref{sec:caHK}) of its data points.

\begin{figure}
\centering
\includegraphics[viewport=1 480 360 705,clip,width=.8\linewidth]{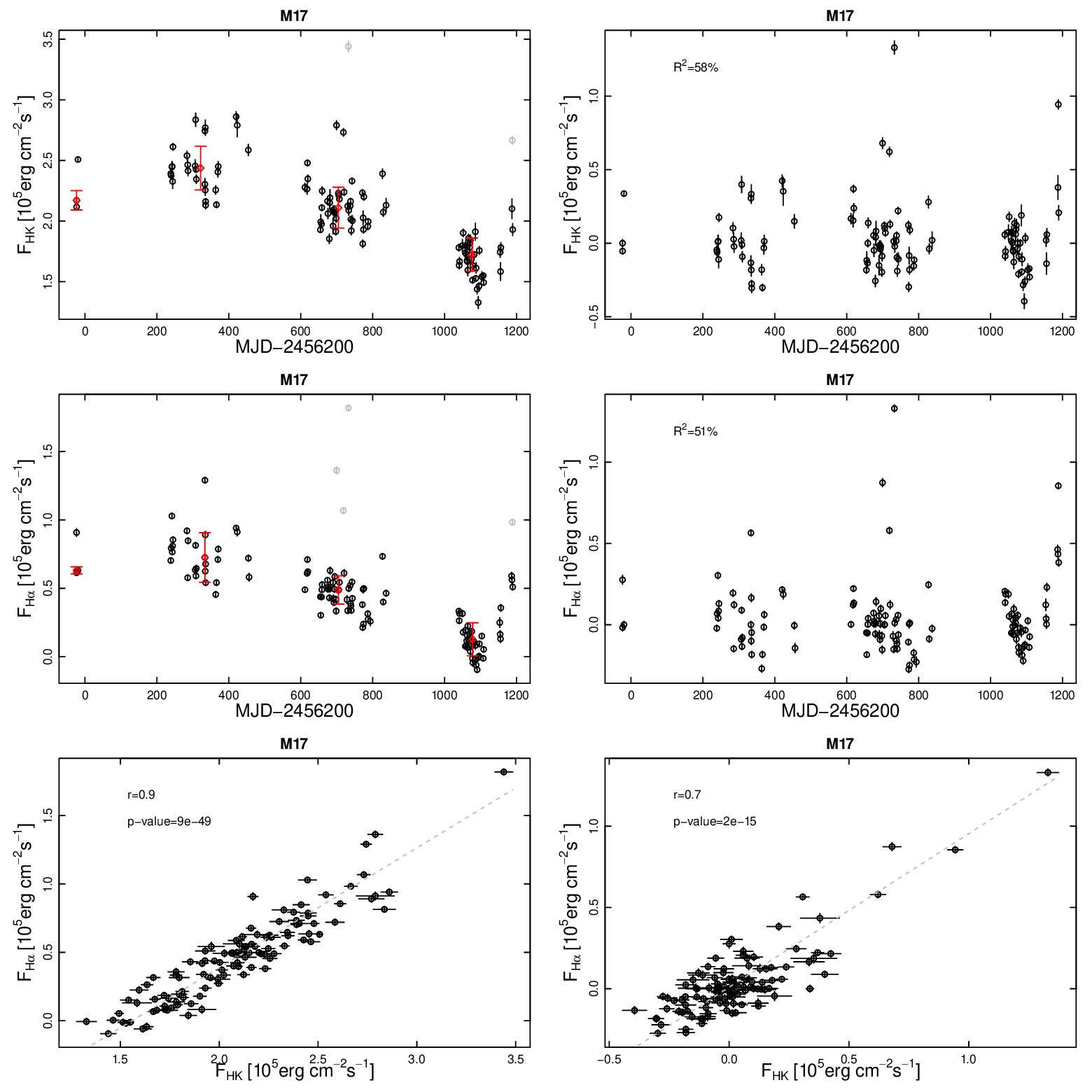}
\caption{Time series of the H\&K emission flux of GJ~4306. For each season, red circles and red error bars represent the median and the MAD of the corresponding sample of measurements. Gray symbols show the rejected outliers.}\label{fig:quadratic}
\end{figure}

First of all, we check if the scatter of each star in the diagram is due to an ongoing long-term activity cycle or it is due to variability on shorter timescales. In this regard, for each star we visually inspect the \ion{Ca}{ii} H\&K and \halpha\ time series, and in some cases we find hints of increasing or decreasing activity over year-long time scales, or even turnarounds (see Fig.~\ref{fig:quadratic} for an example). \citet{Robertson2013} found similar evidence analyzing the $\sim$11-year long series of \halpha\ observations of a sample of 93 M-type dwarfs, and concluded that $\sim$10\% of M dwarfs show either long-term trends or $\gtrsim$1~year activity cycles.

To remove the aforementioned trends, for each season we reject outliers (generally positive outliers likely due to flares) and we compute the median of the remaining measurements. Then, for each season we subtract the corresponding median flux, thus filtering out any $>$1~year long variability. We find that long-term variability is generally able to explain up to 40\% of the scatter in the emission line fluxes. We thus conclude that $\gtrsim$60\% of the variance in the data is due to daily-to-monthly variability.

In support of this conclusion, we find that slopes, correlation coefficients and significances are generally preserved by the correction for long-term trends, which confirms that the displacement of the stars in Fig.~\ref{fig:HalphaHK} is mainly due to correlated shorter-term variability. In the following, we will always refer to the long-term-corrected flux excess.

\begin{figure}
\centering
\includegraphics[viewport=20 50 350 280,clip,width=.8\linewidth]{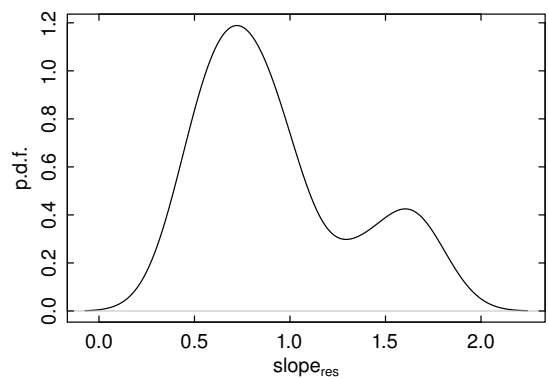}
\caption{Distribution function of the slopes of the $\rm F_{HK}$ vs. $\rm F_{H\alpha}$ relationship with p-value$<$1\%.}\label{fig:slopes_pdf}
\end{figure}

In Fig.~\ref{fig:slopes_pdf} we plot the distribution function of the slopes of the residual $\rm F_{HK}$ vs. $\rm F_{H\alpha}$ relationship for those stars that have slopes with high statistical significance (p-value$<$1\%). This distribution shows that the slope is always positive, i.e.\ $\rm F_{H\alpha}$ is always an increasing function of $\rm F_{HK}$.

Moreover, the distribution function is not symmetric, having a broad right tail. The Anderson-Darling normality test \citep{Anderson1952} supports the non-normality of the sample distribution at the level of $\sim$4 10$^{-3}$. We thus infer that the distribution of the data is not dominated by measurement errors (which would lead to a gaussian distribution function), but reflects an intrinsic characteristic of the analyzed sample.

\begin{figure}
\centering
\includegraphics[viewport=20 50 350 280,clip,width=.8\linewidth]{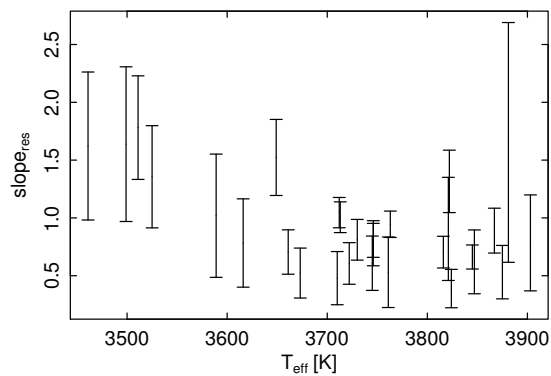}
\caption{Slope of the $\rm F\rm{_{CaII_{HK}}}$ vs. $\rm F\rm{_{H\alpha}}$ relationship as a function of \teff. Only the stars with significant correlation (p-value$<$0.01) have been taken into account.}\label{fig:slopes_parameters}
\end{figure}

We also test the dependence of the slopes of the $\rm F_{HK}$ vs. $\rm F_{H\alpha}$ relationship on $\rm <F_{HK}>$, \teff\ and \met\ respectively, and we find evidence of decreasing steepness of the flux-flux relationship with increasing \teff with a confidence level of $\sim$4\% (Fig.~\ref{fig:slopes_parameters}). \citet{Stelzer2013} found a similar trend analyzing a sample of pre-MS low-mass stars in the 2500--4500 K temperature range. Conversely, Kendall's correlation test cannot reject the hypothesis that the slope of the flux-flux relationship is independent of the average level of activity, or the metallicity of the star. On the other hand, we remark that we cannot claim the independence from the aforementioned parameters, as a lack of statistical dependence may be a consequence of the narrow ranges of the investigated stellar parameters.

We compare these results with the model provided by \citet{Meunier2009}, who explain the $\rm F_{HK}$ vs. $\rm F_{H\alpha}$ relationship for the Sun. In their work, the authors use spatially resolved observations of the solar surface covering 1.5 activity cycles. They find that the presence of dark filaments affects the correlation between $\rm F_{HK}$ and $\rm F_{H\alpha}$ on short (daily-weekly) timescales, sometimes leading to negative correlations depending on the surface coverage of plages and filaments. Analogous phenomena may affect the distribution in Fig.~\ref{fig:slopes_pdf} leading to a non-gaussian shape, and may also explain why we obtain negative slopes in some cases, even though with low statistical significance (see Fig.~\ref{fig:HalphaHK}, gray lines in the right panel). In particular, in their model, larger values for the slopes correspond to a scenario in which the chromospheric \halpha\ emission is less affected by the absorption of cold filaments compared with the rest of the sample. For a fixed configuration of plages seen in the \ion{Ca}{ii} H\&K lines on the Sun, the slope of the flux-flux relationship depends on the contrast of the \halpha\ emission in the plages and the absorption by filaments, and their respective filling factors. For a fixed contrast of \halpha\ plages on the solar disk, the slope of the $\rm F_{HK}$ vs. $\rm F_{H\alpha}$ relationship tends to decrease as the contrast of the filaments increases. This is explained by the fact that solar filaments are optically thicker in the \halpha\ as compared to in the \ion{Ca}{ii} H\&K lines. Thus, the increase of the slope with decreasing \teff\ (Fig.~\ref{fig:slopes_parameters}) may indicate that the \halpha\ absorption in equivalent active regions decreases toward later M types. Analyzing the Balmer decrements of the same sample of stars we independently draw the same conclusion \citep{Maldonado2016}, consistently with similar previous studies \citep{Bochanski2007,Stelzer2013}.

\section{Chromospheric variability}\label{sec:variability}

\subsection{Variability vs. activity vs. age}\label{sec:varactage}
%gsAnalyzeVariabilitySeasonalMedian.r

\begin{figure}
\centering
\includegraphics[width=\linewidth]{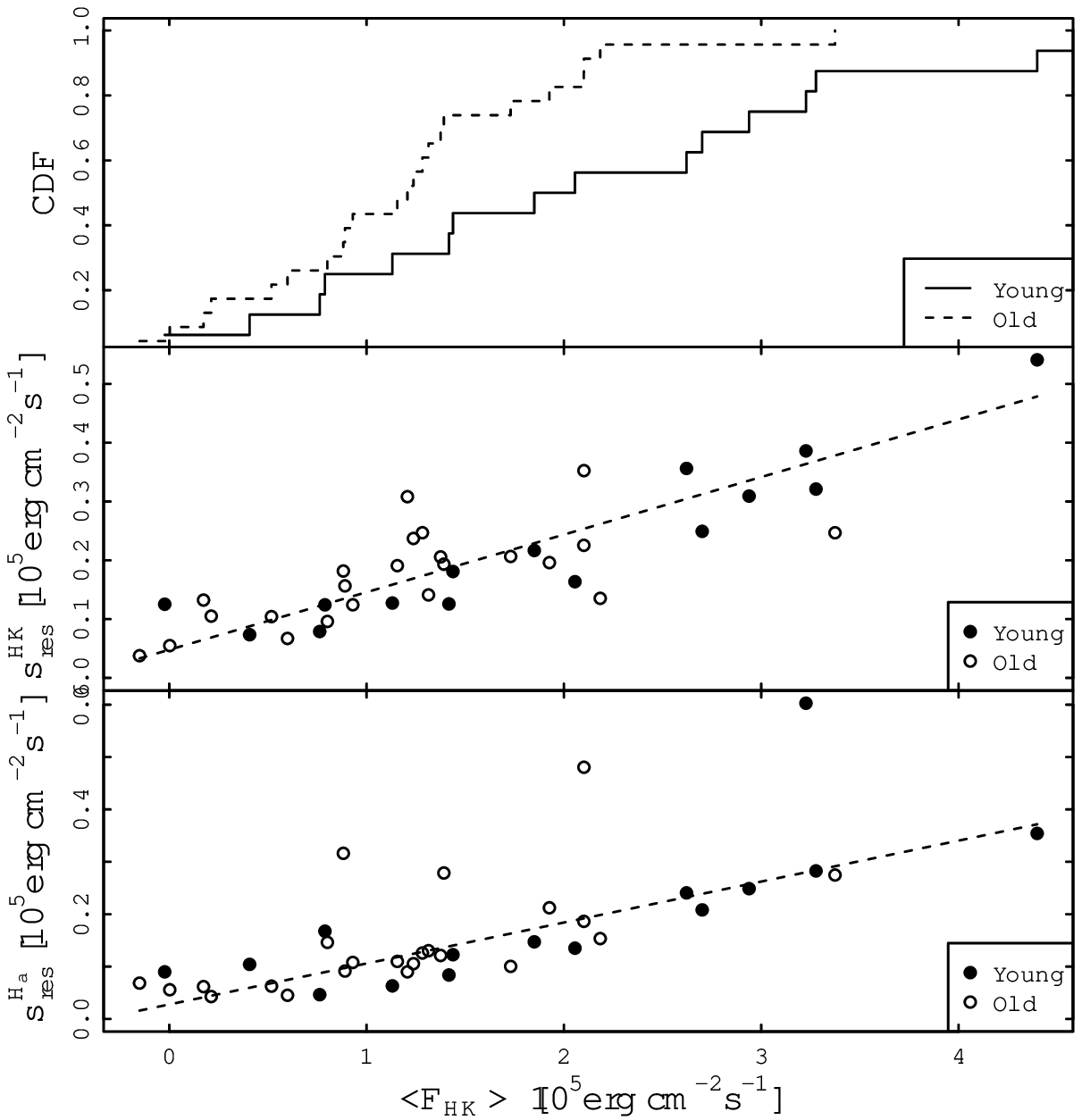}
\caption{\textit{Top panel - } Empirical cumulative distribution function of the $<F_{HK}>$ measurement for young (solid line) and old (dashed line) stars, following the classification of \citet{Maldonado2016} based on kinematics. \textit{Central panel - } Standard deviation $\sigma^{\rm HK}_{\rm res}$ of the H\&K measurements corrected for the quadratic trend (Sect.~\ref{sec:HKHalpha}) as a function of the average level of activity $<F_{HK}>$. The black dashes represent the best-fit line, the best-fit coefficients are reported in Table~\ref{tab:coeffSigma}. Filled and open circles mark young and old stars respectively. \textit{Bottom panel - } Same as in the middle panel, for the residual \halpha\ line.}\label{fig:sigma}
\end{figure}

In Fig.~\ref{fig:sigma} we plot the standard deviation of the residual \ion{Ca}{ii} H\&K and \halpha\ flux measurements (see Sect.~\ref{sec:HKHalpha}) of each program star ($\sigma^{\rm HK}_{\rm res}$ and $\sigma^{\rm H\alpha}_{\rm res}$ respectively) as a function of the median level of \ion{Ca}{ii} H\&K flux, assumed to monotonically increase with the level of magnetic activity (see Sect.~\ref{sec:fluxflux}). In these diagrams we find a tight correlation (supported by the Spearman's correlation test) between variability and median excess flux. The linear best fits suggest that (1)  the fractional variability with respect to the average \ion{Ca}{ii} H\&K flux excess (which is represented by the slope of the linear fit) is roughly 10\% (2) even at the minimum level of activity there is a residual variability of $\sim\rm 0.5~10^{-5} erg\ cm^{-2}\ s^{-1}$, which is different from zero at the $\sim$2$\sigma$ level for the \ion{Ca}{ii} H\&K lines (see Table~\ref{tab:coeffSigma}). In the case this is a real effect rather than a bias introduced by our analysis, it is an evidence that even at minimum activity there is some variability.

We also separate the stars younger and older than $\sim$650~Myr as classified in \citet{Maldonado2016} based on their kinematics (Table~\ref{tab:stars}). Despite the fact that kinematics provide only lower limits to stellar ages \citep[][and references therein]{Maldonado2010}, with this classification we find evidence that the two populations of stars show significantly different distributions in terms of \ion{Ca}{ii} H\&K emission flux. The Kolmogorov-Smirnov test on the $\rm<F_{HK}>$ we measure confirms this result (top panel in Fig.~\ref{fig:sigma}), indicating that younger stars have higher \ion{Ca}{ii} H\&K fluxes. The same test on $\rm<F_{H\alpha}>$ is less conclusive. This is likely due to the fact that the \halpha\ is seen in absorption for a range of $\rm<F_{HK}>$, i.e. the \halpha\ is not a good diagnostic at low and intermediate activity levels (see Sect.~\ref{sec:HKHalpha}).

The Kolmogorov-Smirnov test on the $\sigma^{\rm HK}_{\rm res}$ and $\sigma^{\rm H\alpha}_{\rm res}$ does not provide any firm indication that the amount of variability is age-dependent, i.e. variability is not a good age estimator.

\begin{table}
 \begin{center}
\caption{Best-fit models of the form $y=q+m\cdot x$ shown in Fig.~\ref{fig:sigma}.}\label{tab:coeffSigma}
  \begin{tabular}{cccc}
 \hline\hline
y & x & q$\rm\pm\sigma_q$ & m$\rm\pm\sigma_m$ \\
$\rm [erg\ cm^{-2}\ s^{-1}]$ & $\rm [erg\ cm^{-2}\ s^{-1}]$ & $\rm [erg\ cm^{-2}\ s^{-1}]$ & \\
\hline
$\sigma^{\rm HK}_{\rm res}$ & $\rm<F_{HK}>$ & 0.05$\pm$0.03 & 0.10$\pm$0.02\\
$\sigma^{\rm H\alpha}_{\rm res}$ & $\rm<F_{HK}>$ & 0.03$\pm$0.03 & 0.08$\pm$0.02\\
\hline
  \end{tabular}
 \end{center}
\end{table}

\begin{figure}
\centering
\includegraphics[width=\linewidth]{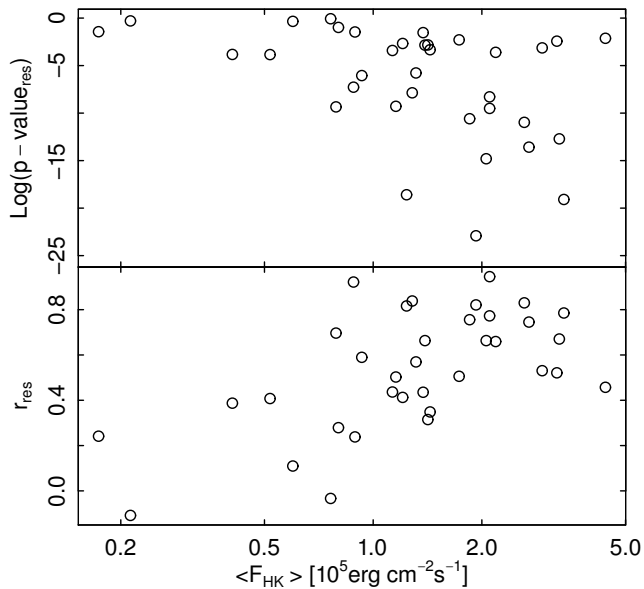}
\caption{p-values (top panel) and Pearson's correlation coefficient (bottom panel) of the best-fits between \halpha\ and \ion{Ca}{ii} H\&K fluxes shown in Fig.~\ref{fig:HalphaHK} vs. the average level of activity $\rm<F_{HK}>$.}\label{fig:slopes_r}
\end{figure}

One may argue that the increase of variability in the \ion{Ca}{ii} H\&K and \halpha\ residual emission fluxes may lower the degree of correlation in the flux-flux relationship. On the contrary, we find that larger variability generally corresponds to higher levels of \ion{Ca}{ii} H\&K activity (Fig.~\ref{fig:sigma}, middle panel), which in turn are associated with stronger correlation (Pearson's coefficient $r_{res}$ closer to 1, smaller p-values$\rm_{res}$) of the flux-flux relationship (Fig.~\ref{fig:slopes_r}). Conversely, the quietest stars tend to show high p-values$\rm_{res}$, and in general have lower or negative correlation coefficients r$\rm_{res}$. The same results were also found by \citet{Gomes2011} analyzing the long-term activity of a sample of M dwarfs. This supports the assumption that line emission in the \ion{Ca}{ii} H\&K doublet and the \halpha\ line are triggered by spatially correlated ARs.

\subsection{Variability timescales}
%gsPooledVariance.r

To analyze the typical timescales of the variability, we perform a time series analysis using the \lq\lq Pooled Variance\rq\rq\ (PV) approach described by \citet{Donahue1,Donahue2}.

The PV is a measure of the average variance in the data over the timescale $\tau$. By construction, the PV computed at $\tau$ is expected to be the combination of contributions from a variety of sources, instrumental and/or astrophysical, with timescales shorter than $\tau$. When $\tau$ increases, the PV remains constant until the effects from processes with longer timescales become more noticeable, in which case the PV increases with $\tau$. The PV approach is thus suited for time series containing multiple periodic signals with different amplitudes and phases, and is indicated to study the timescales of variability induced by stellar rotation and the life cycle of ARs.

\begin{figure*}
\centering
\includegraphics[width=.33\linewidth,viewport=1 217 355 410,clip]{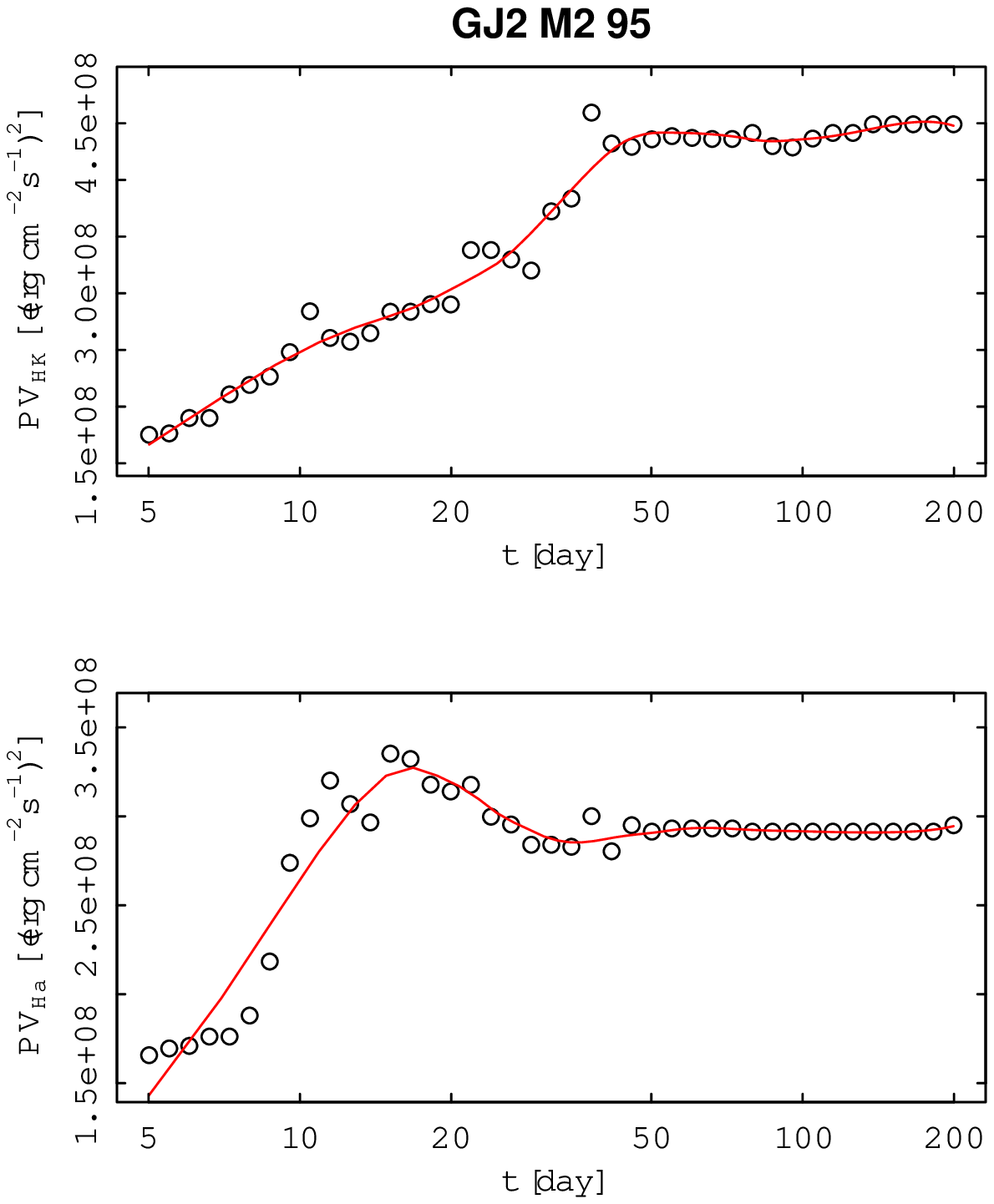}%
\includegraphics[width=.33\linewidth,viewport=1 1 355 194,clip]{pv2.eps}%
\includegraphics[width=.33\linewidth,viewport=1 1 355 191,clip]{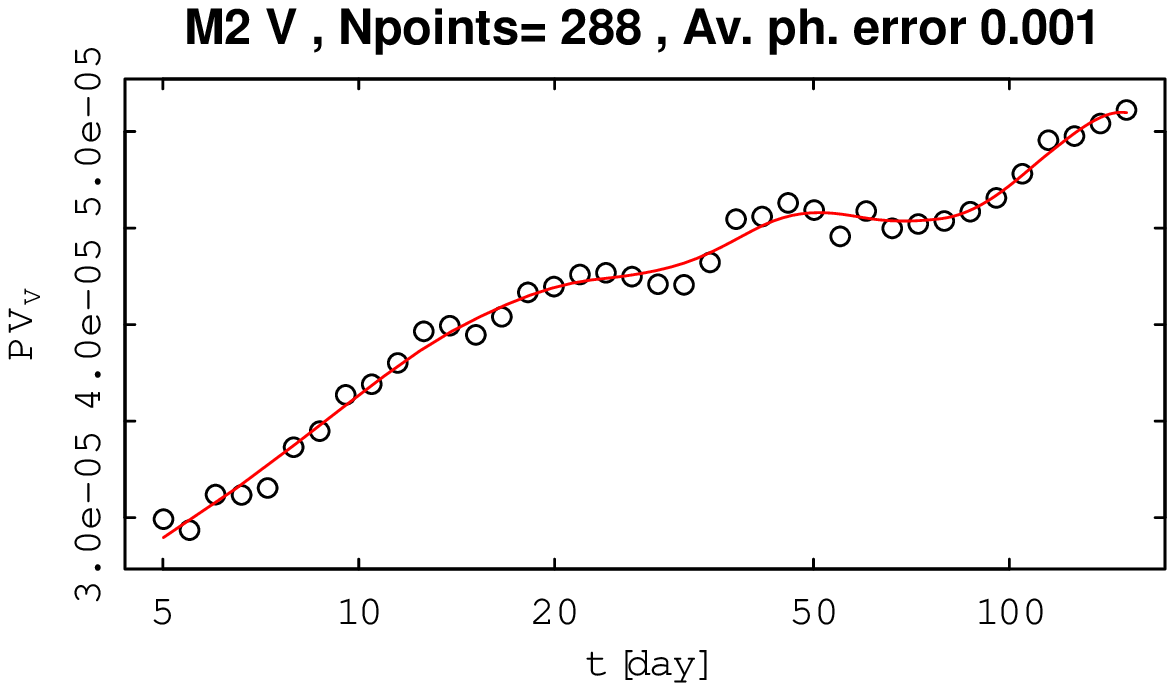}%
\caption{PV diagrams of GJ~2. From left to right, the PV diagrams of $\rm F_{HK}$, $\rm F_{Halpha}$ and V photometric band are shown respectively. The red line is a smoothing function to ease the reading of the graphs.
}\label{fig:pv2}
\end{figure*}

\begin{figure*}
\centering
\includegraphics[width=.33\linewidth,viewport=1 217 355 410,clip]{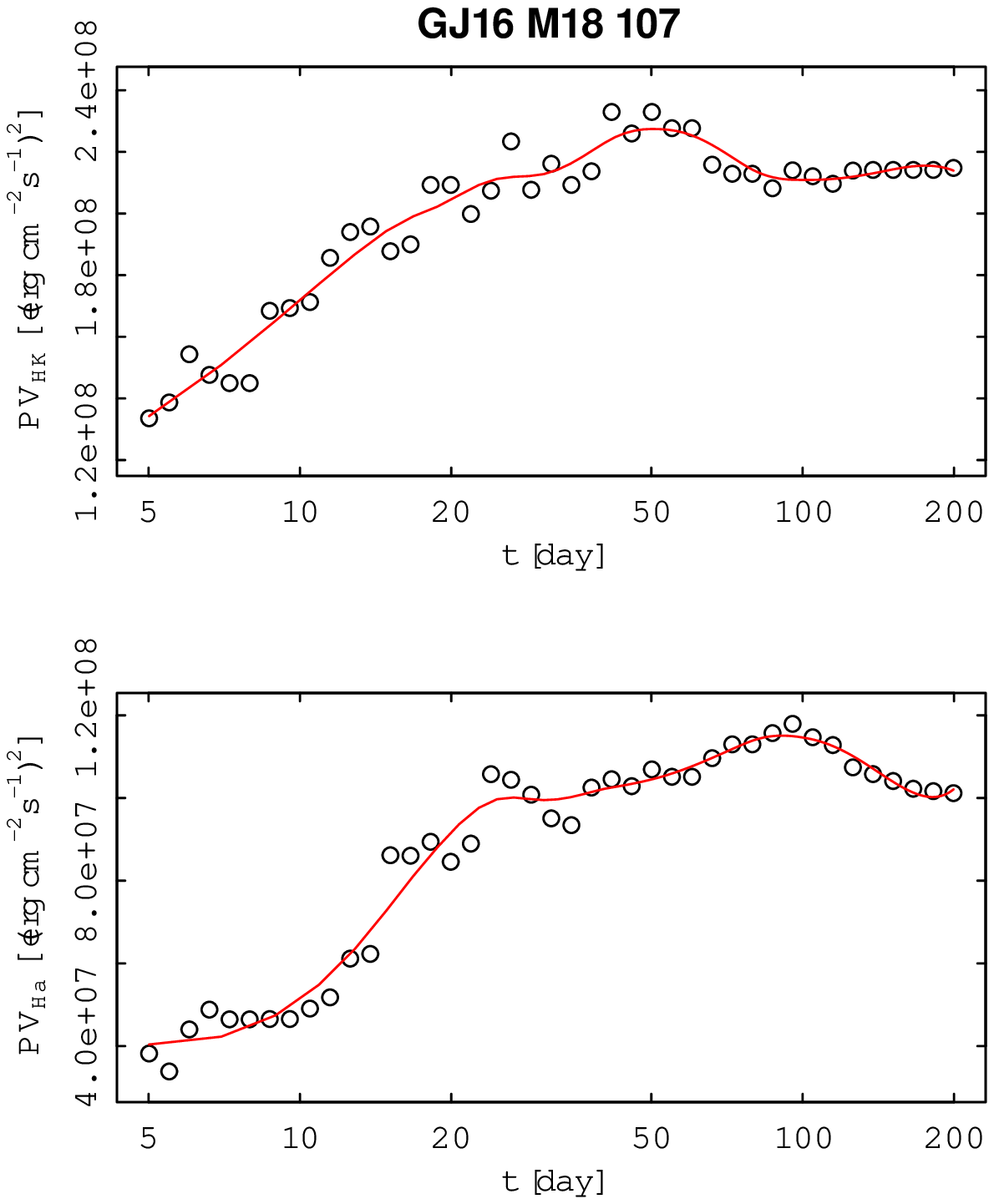}%
\includegraphics[width=.33\linewidth,viewport=1 1 355 196,clip]{pv16.eps}%
\includegraphics[width=.33\linewidth,viewport=1 1 355 191,clip]{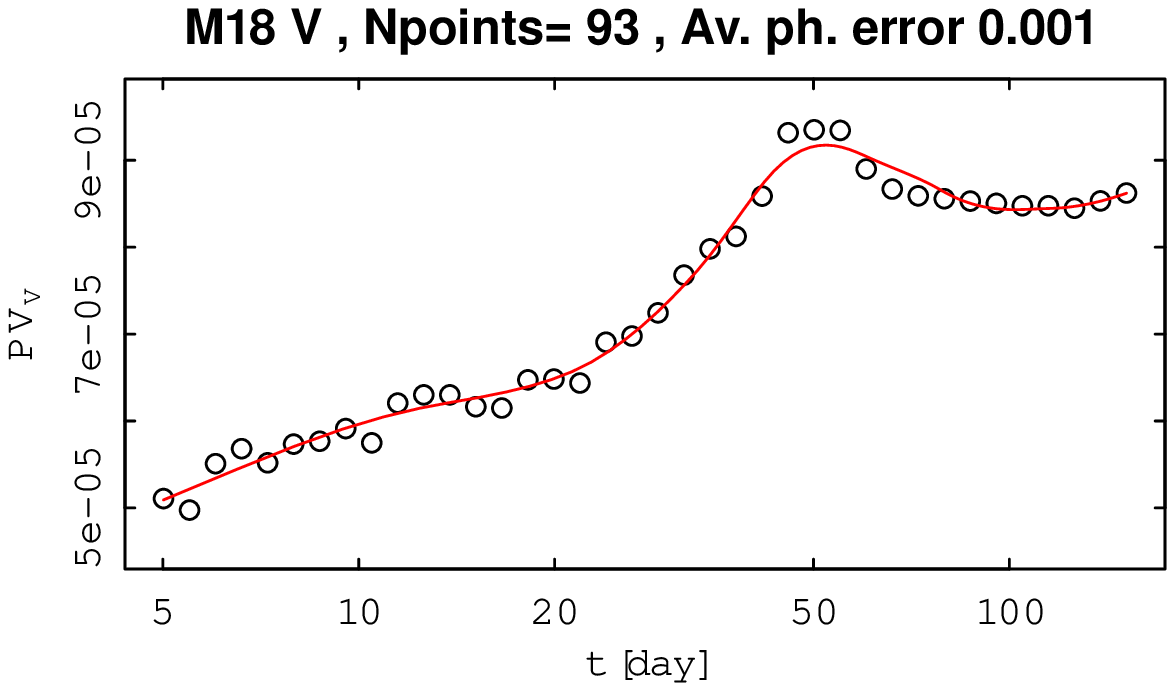}%
\caption{Same as in Fig.~\ref{fig:pv2}, for GJ~16.}\label{fig:pv16}
\end{figure*}

\begin{figure*}
\centering
\includegraphics[width=.33\linewidth,viewport=1 217 355 410,clip]{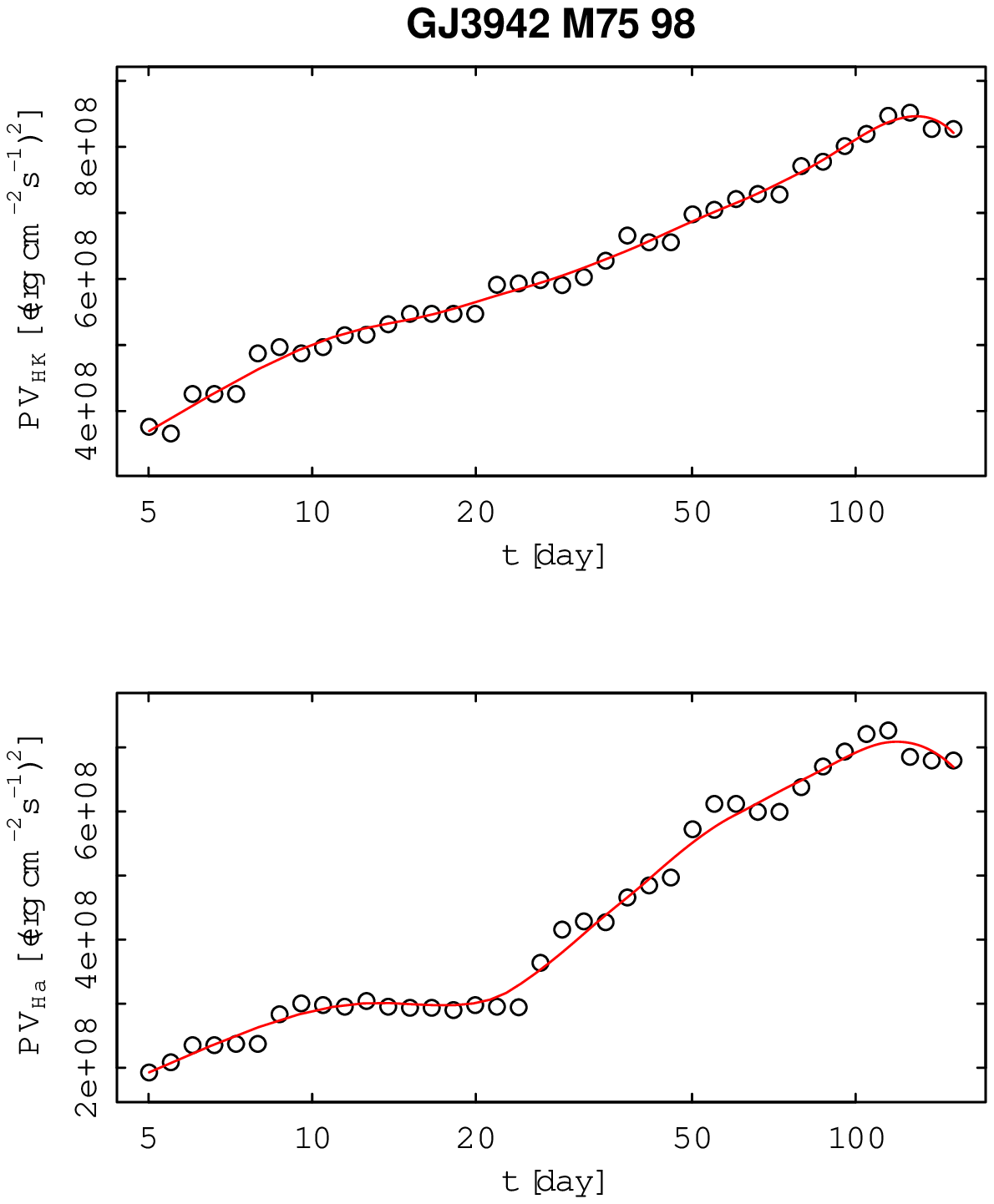}%
\includegraphics[width=.33\linewidth,viewport=1 1 355 196,clip]{pv3942.eps}%
\includegraphics[width=.33\linewidth,viewport=1 1 355 191,clip]{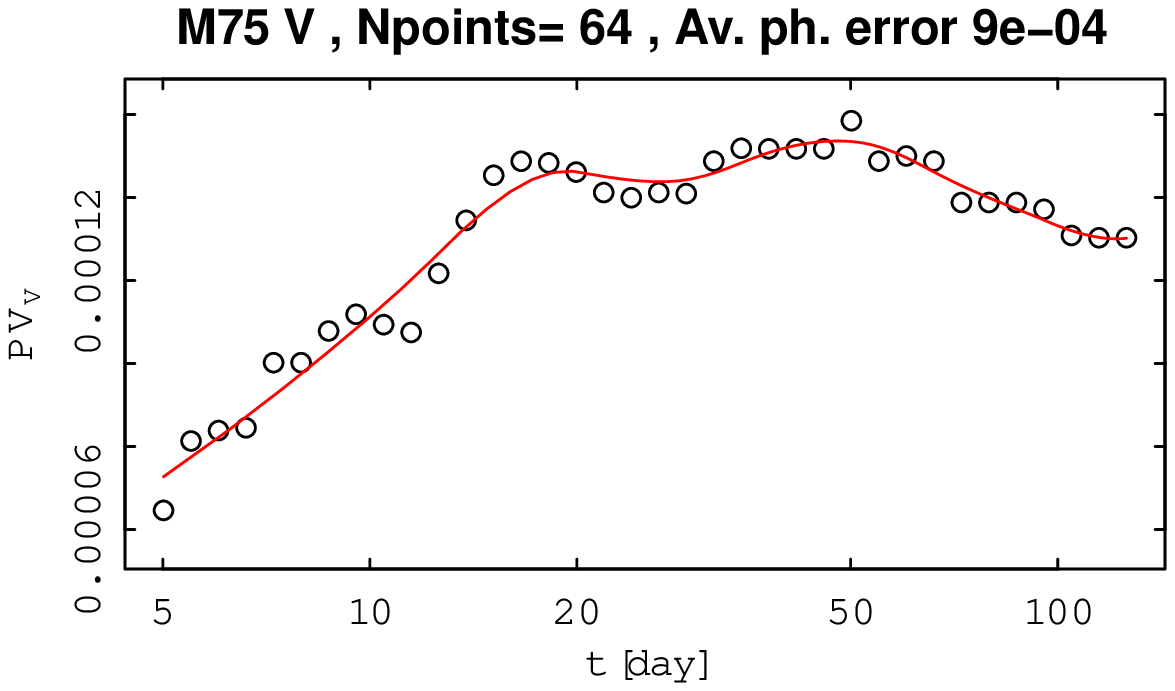}%
\caption{Same as in Fig.~\ref{fig:pv2}, for GJ~3942.}\label{fig:pv3942}
\end{figure*}

\begin{figure*}
\centering
\includegraphics[width=.33\linewidth,viewport=1 217 355 410,clip]{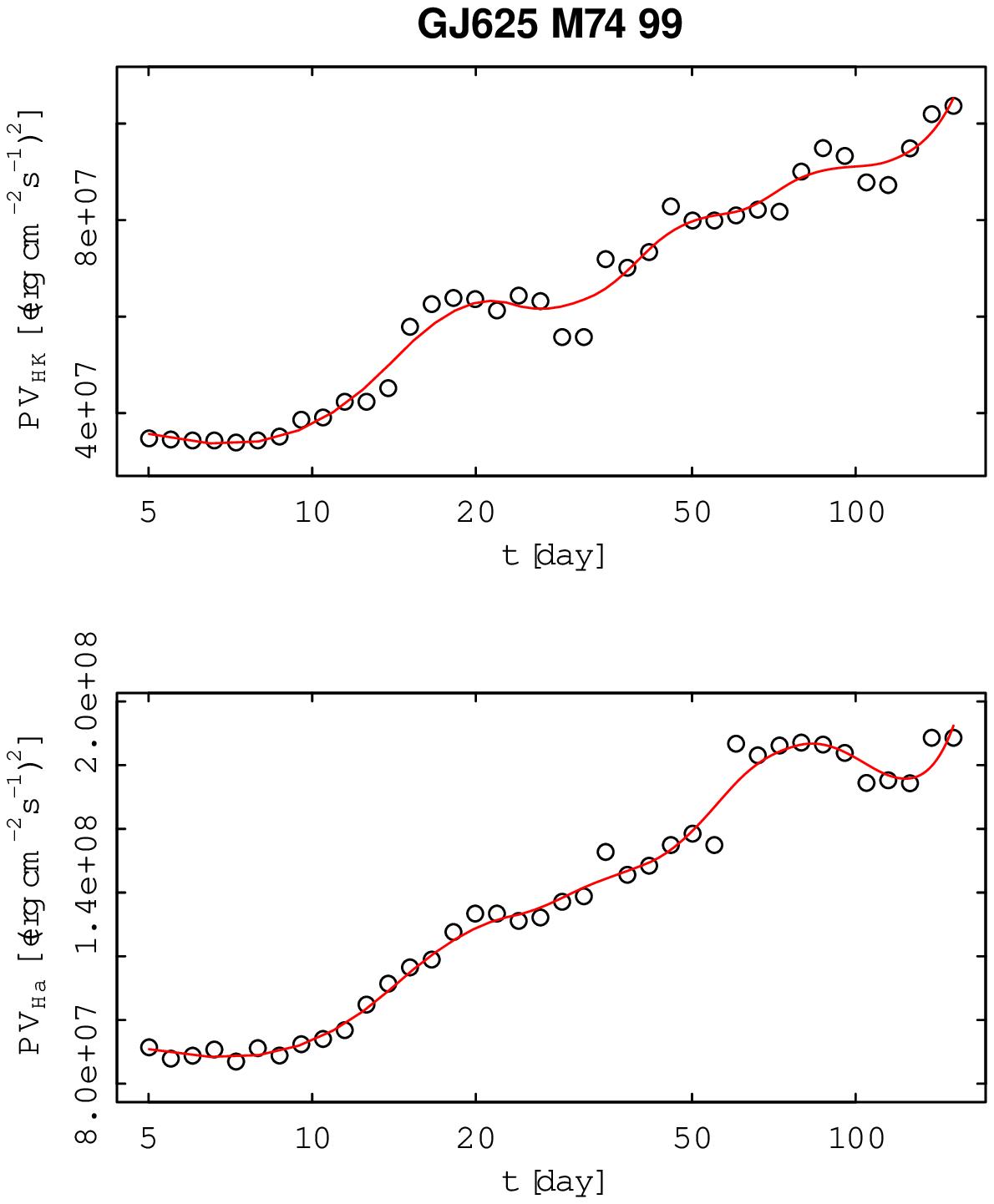}%
\includegraphics[width=.33\linewidth,viewport=1 1 355 196,clip]{pv625.eps}%
\includegraphics[width=.33\linewidth,viewport=1 1 355 191,clip]{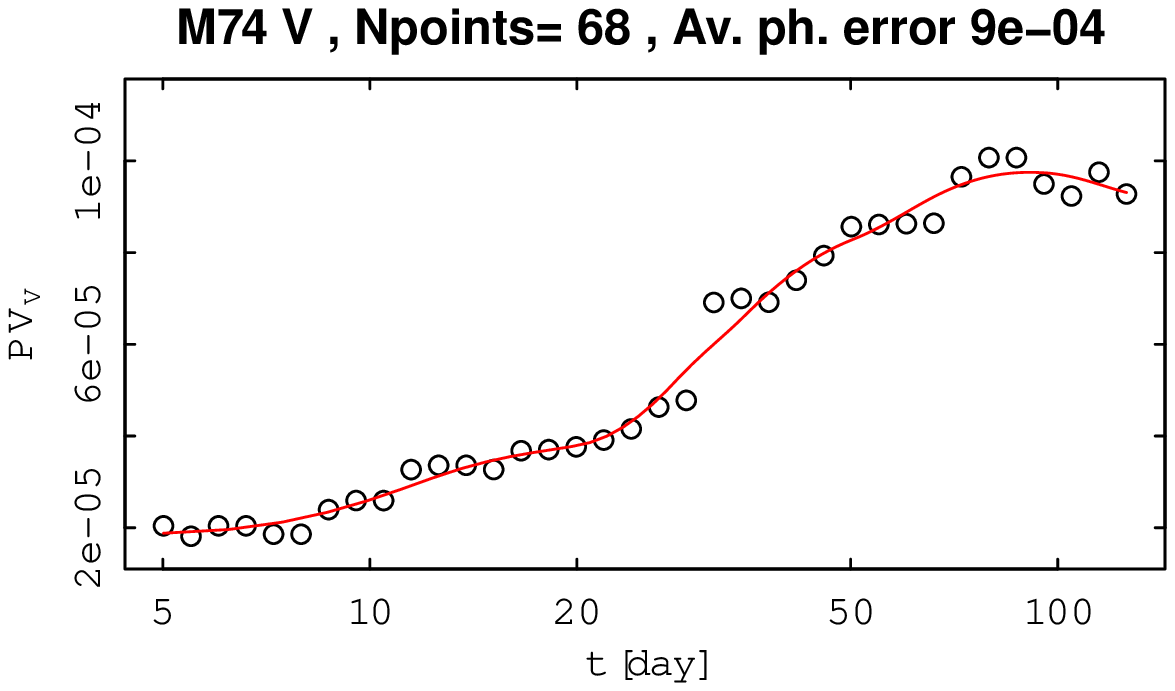}%
\caption{Same as in Fig.~\ref{fig:pv2}, for GJ~625.}\label{fig:pv625}
\end{figure*}

\begin{figure*}
\centering
\includegraphics[width=.33\linewidth,viewport=1 217 355 410,clip]{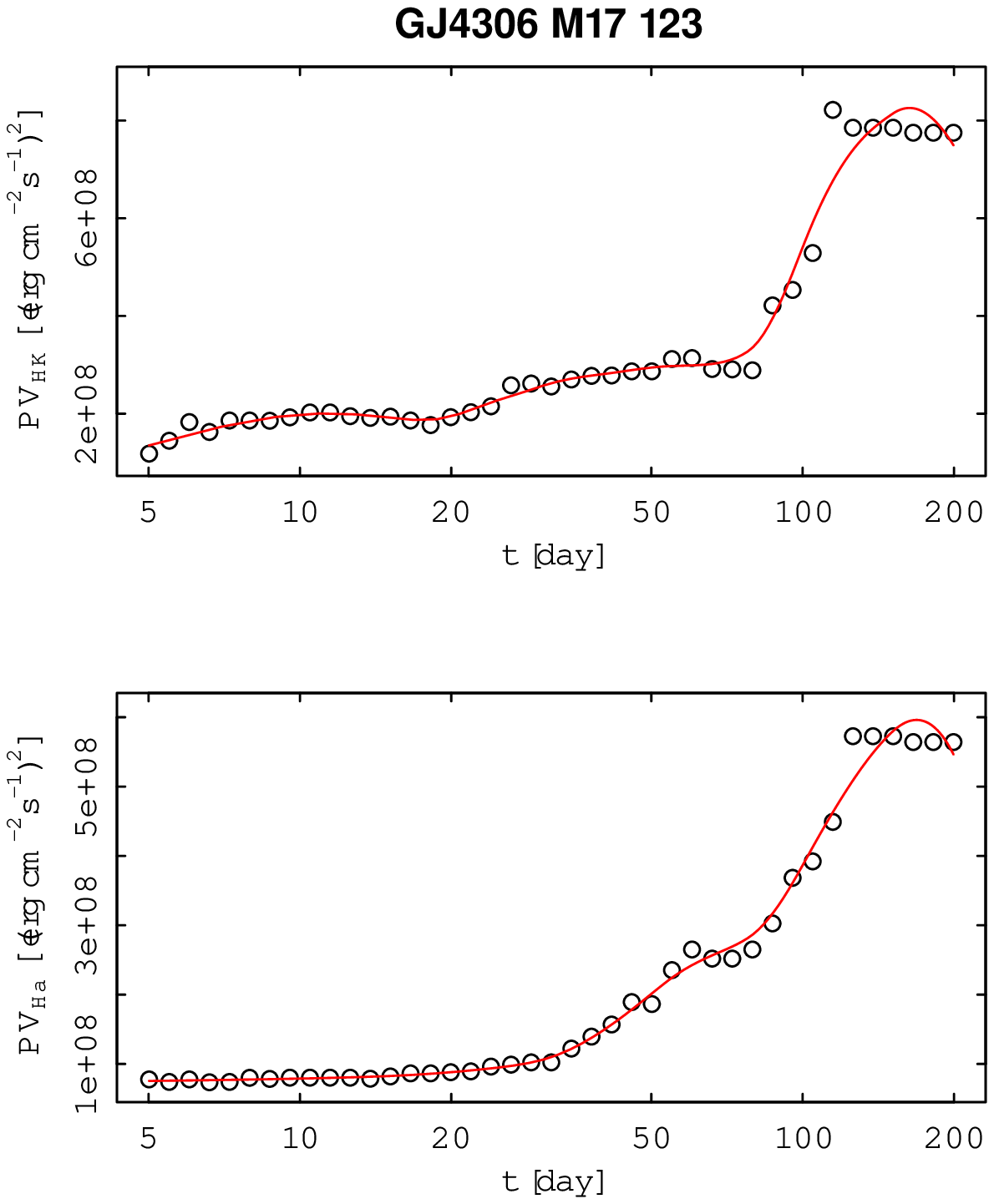}%
\includegraphics[width=.33\linewidth,viewport=1 1 355 196,clip]{pv4306.eps}%
\includegraphics[width=.33\linewidth,viewport=1 1 355 191,clip]{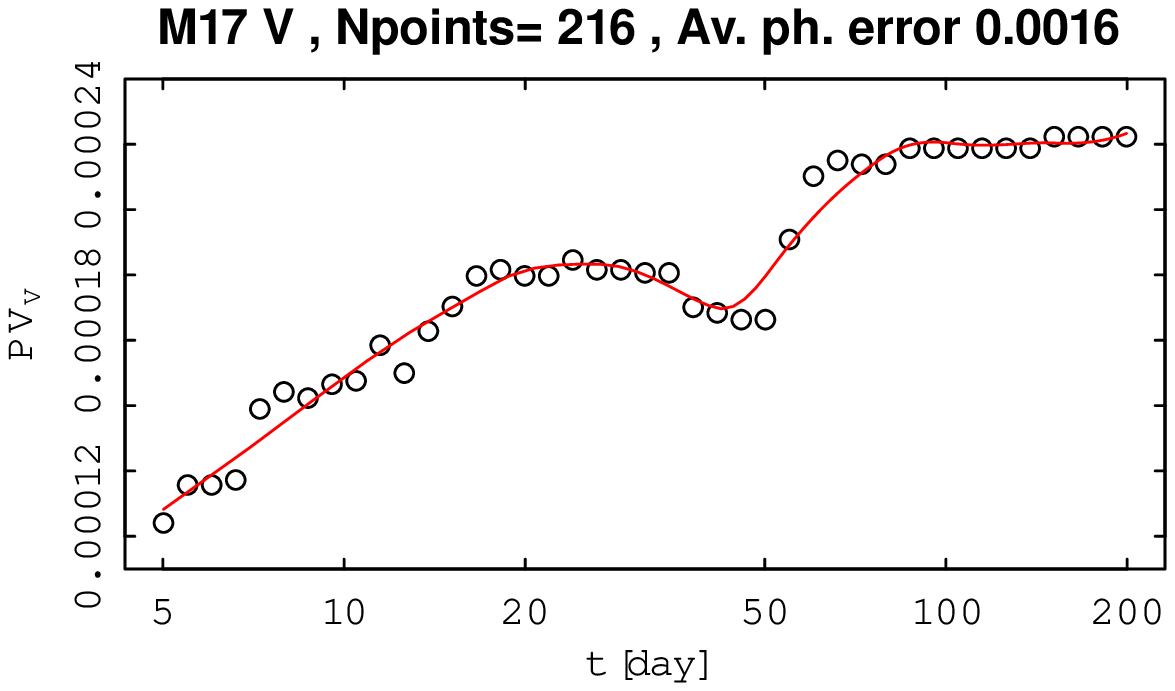}%
\caption{Same as in Fig.~\ref{fig:pv2}, for GJ~4306.}\label{fig:pv4306}
\end{figure*}

\begin{figure*}
\centering
\includegraphics[width=.33\linewidth,viewport=1 217 355 410,clip]{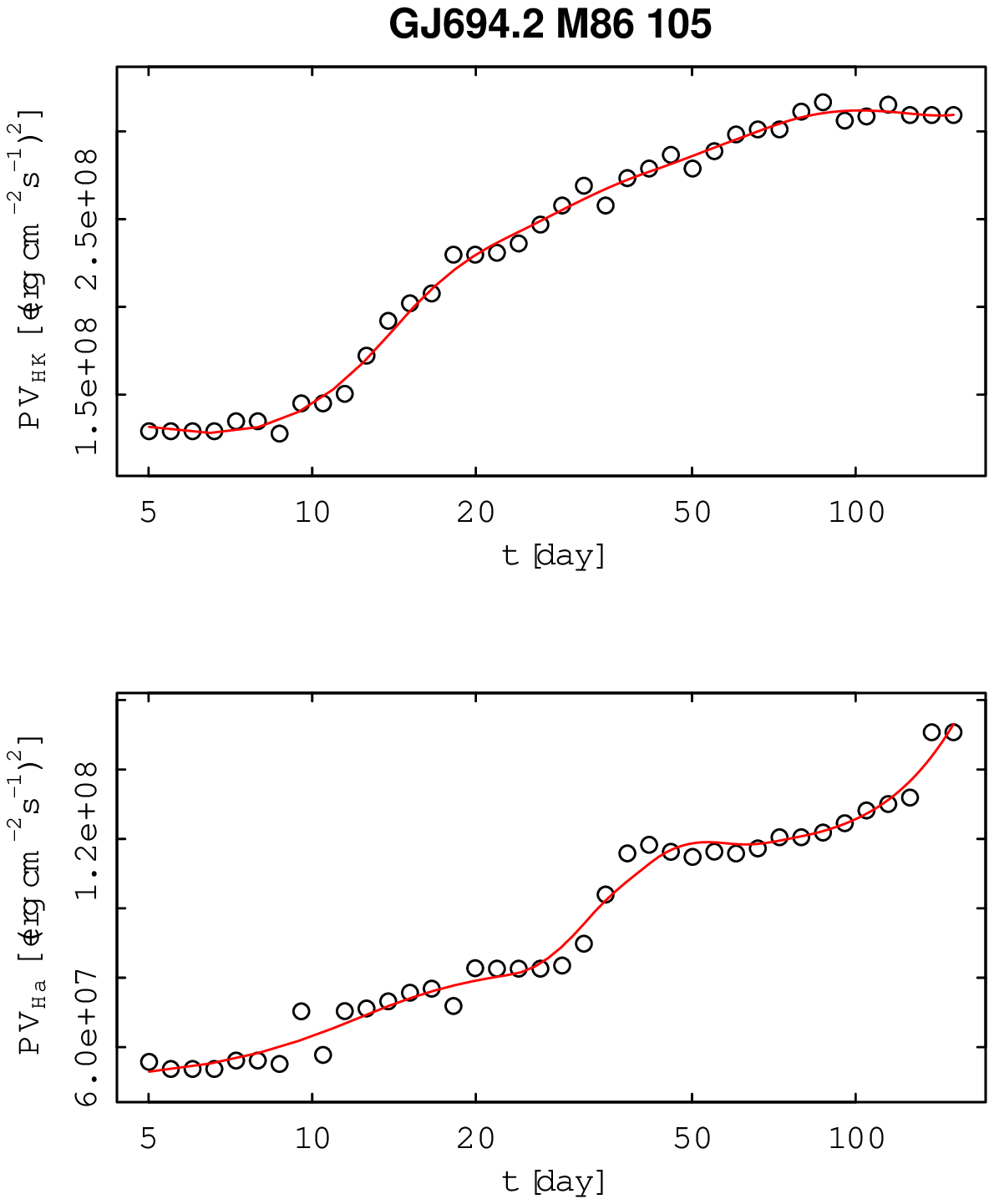}%
\includegraphics[width=.33\linewidth,viewport=1 1 355 196,clip]{pv694.eps}%
\includegraphics[width=.33\linewidth,viewport=1 1 355 191,clip]{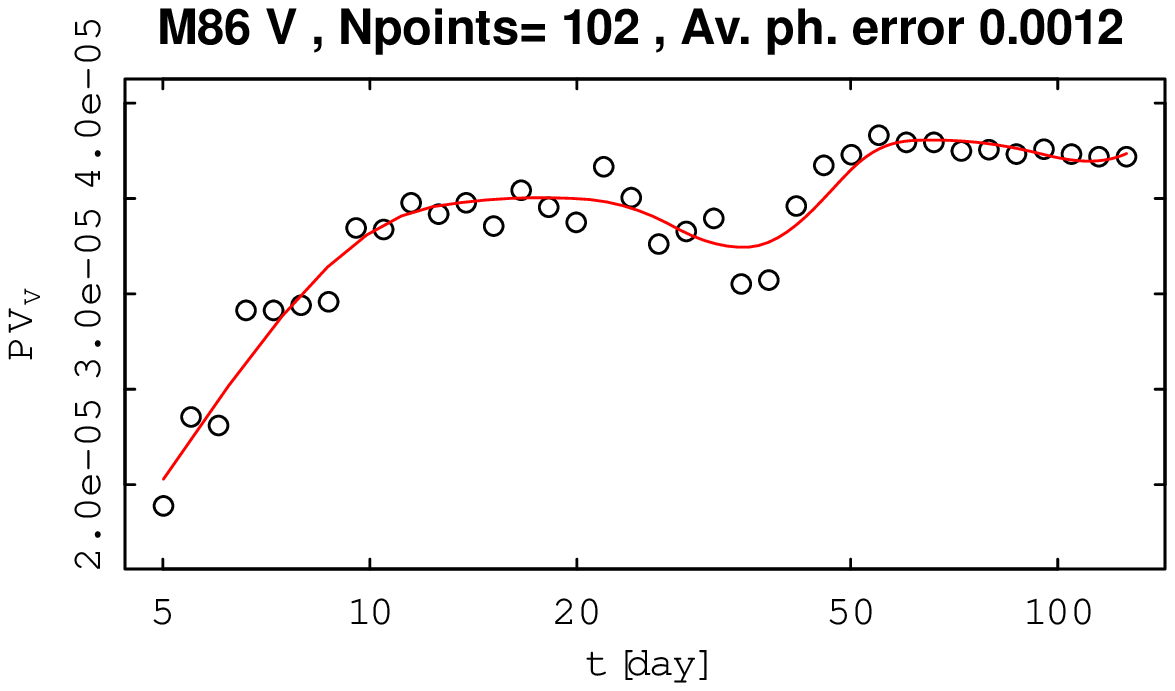}%
\caption{Same as in Fig.~\ref{fig:pv2}, for GJ~694.}\label{fig:pv694}
\end{figure*}

\begin{figure*}
\centering
\includegraphics[width=.33\linewidth,viewport=1 217 355 410,clip]{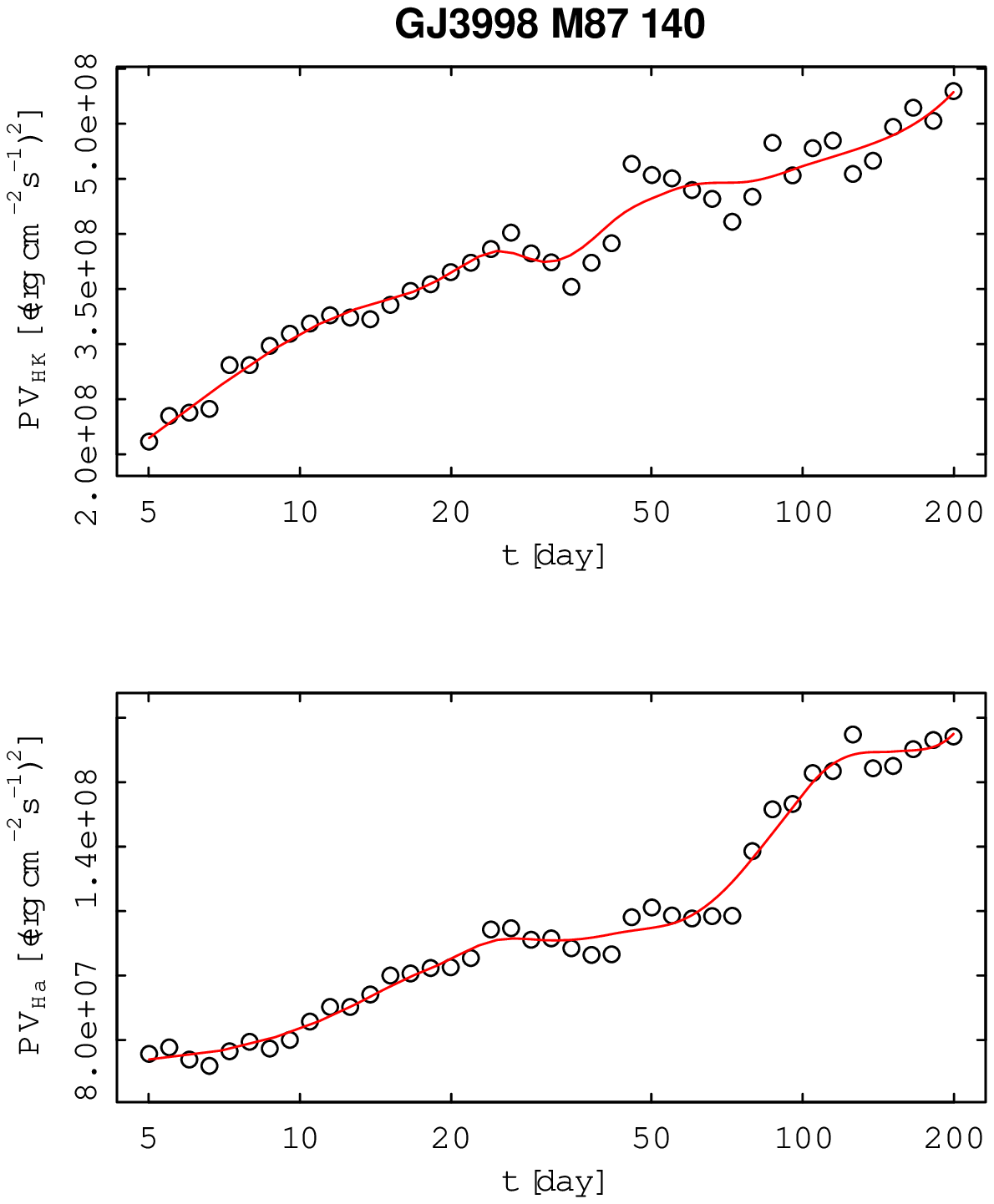}%
\includegraphics[width=.33\linewidth,viewport=1 1 355 196,clip]{pv3998.eps}%
\includegraphics[width=.33\linewidth,viewport=1 1 355 191,clip]{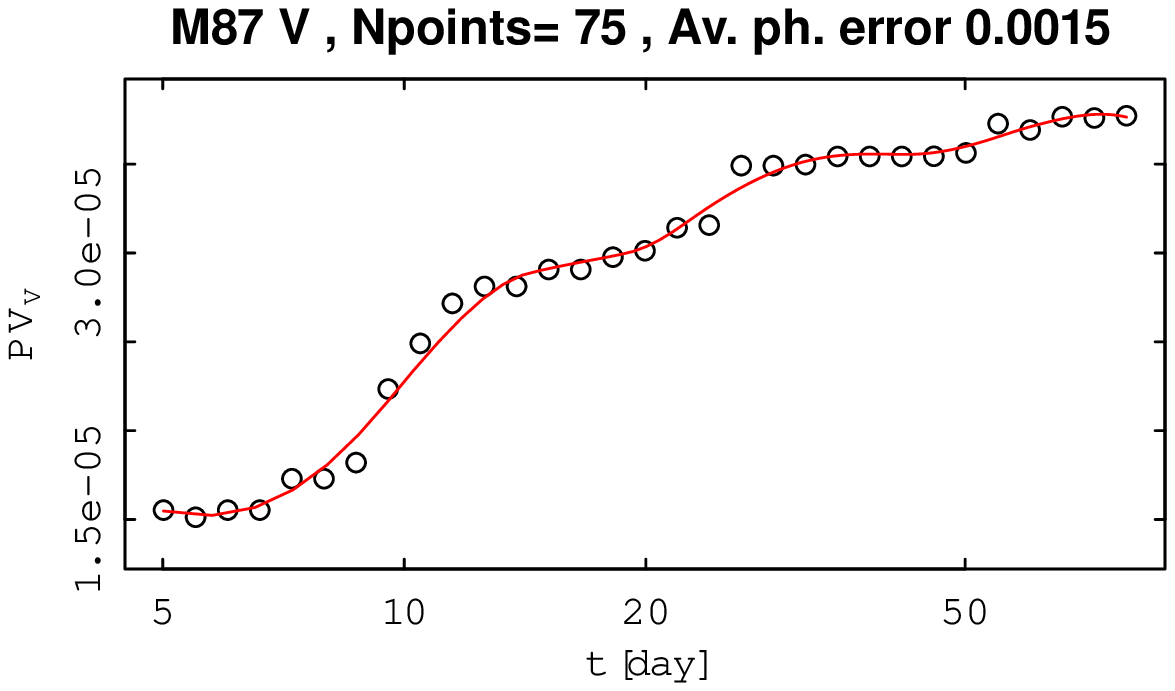}%
\caption{Same as in Fig.~\ref{fig:pv2}, for GJ~3998.}\label{fig:pv3998}
\end{figure*}

\begin{figure*}
\centering
\includegraphics[width=.33\linewidth,viewport=1 217 355 410,clip]{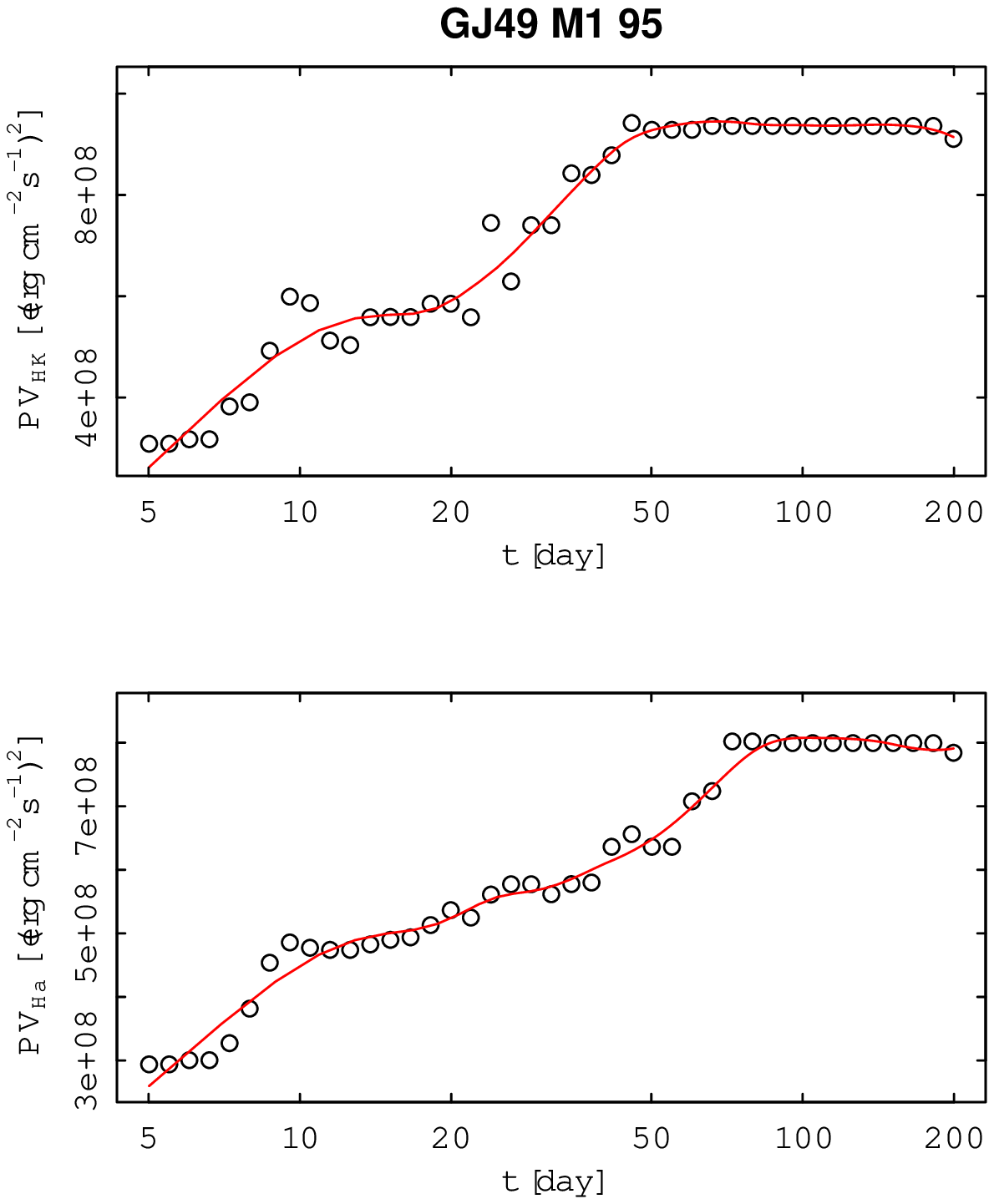}%
\includegraphics[width=.33\linewidth,viewport=1 1 355 196,clip]{pv49.eps}%
\caption{Same as in Fig.~\ref{fig:pv2}, for GJ~49. The V band monitoring of this star is not long enough to allow the PV analysis.}\label{fig:pv49}
\end{figure*}

The PV computation needs a large number of data to be robust \citep{Lanza2004}. In 3.5 years of observation we have collected up to 100 spectra for a single star. Since this is not a particularly abundant data set, we apply the PV analysis only to the 8 stars with more than 90 observations, namely GJ~2, GJ~16, GJ~3942, GJ~625, GJ~4306, GJ~694.2, GJ~3998 and GJ~49, and show their PVs in Figs.~\ref{fig:pv2} to \ref{fig:pv49}.

In the computation of the PV diagrams, we consider the \ion{Ca}{ii} H\&K and \halpha\ flux excesses corrected for the long-term trend (Sect.~\ref{sec:HKHalpha}), hence we are confident that any contribution from long-term activity cycles is reduced to a minimum. We use the algorithm described in \citet{Donahue1}, with the only difference that we use the median and the MAD instead of the mean and the standard deviation respectively, the former being more robust to the presence of outliers especially for small sample sizes.

We complement the spectroscopic data with the V-band photometric monitoring of the same sample of stars, carried out at INAF-Catania Astrophysical Observatory in the same epochs of the spectroscopic monitoring using a 80~cm telescope. The complete and detailed analysis of the photometric monitoring will be discussed in a future paper.

In general, for each star we find that the \ion{Ca}{ii} H\&K, \halpha\ and V band PV diagrams look alike. The PV diagrams generally increase at small timescales $\tau$, they reach a plateau at roughly $\tau\simeq$10-40 days, and then increase to level off again at $\tau\gtrsim$50 days.

Our interpretation of these diagrams is that the first plateau corresponds to the stellar rotation period. This is confirmed by a complementary Lomb-Scargle based analysis\footnote{The full details of the methodology are described in \citet{Suarez2015}, where the authors analyze the rotation of a sample of 48 late F-type to mid-M dwarf stars.} of the \ion{Ca}{ii} H\&K and \halpha\ fluxes we are carrying out to measure the rotation period of the stars \citep[see][]{rotation}. The increase at longer $\tau$ indicates that there is an additional source of variance in the data with timescales longer than the rotation period. The PV then levels off at $\tau\sim$50--60 days, suggesting that the new source of variance typically has time scales of a few stellar rotations. We exclude that the second flattening is due to cyclic activity for two reasons: as a matter of fact, dynamo cycles of M dwarfs act on year-long timescales, as indicated by \citet{Robertson2013}, and their effects should be greatly reduced due to the long-term correction we applied in Sect.~\ref{sec:HKHalpha}. We thus infer that this source of variance is likely related to the growth and decay of chromospheric ARs, as suggested also by \citet{Donahue1,Donahue2}. Other studies have found that M dwarfs may show variability on time scales of the order of 100 days \citep{Davenport2015,Robertson2015,Newton2016}, proposing that it is related to the lifecycle of ARs.

%__________________________________________________________________

\section{Summary}

In this study we analyze the spectroscopy database collected in the framework of the HADES project, focusing on the characterization of the variability of \ion{Ca}{ii} H\&K and \halpha\ lines as chromospheric diagnostics. The database consists of the intensive spectroscopic monitoring performed with the HARPS-N@TNG spectrograph for a sample of 71 low-activity early-type M dwarfs, to search for planets.

We focus on the simultaneous analysis of the flux excess emitted in the \ion{Ca}{ii} H\&K and \halpha\ lines. For this purpose, we developed a technique to correct the spectra for instrumental and atmospheric effects by means of synthetic spectra. This leads to the calibration of the spectra to a common absolute flux scale, which enables the measurement of line flux excesses in units of flux at the stellar surface.

Our measurements show that the \ion{Ca}{ii} H and K flux excesses are strongly linearly correlated, consistent with previous results \citep[see][and references therein]{Martinez2011}. When comparing the \ion{Ca}{ii} H\&K with the \halpha\ chromospheric line flux we find significantly more scatter, larger than the measurement uncertainties. In the $\rm F_{HK}$ vs. $\rm F_{H\alpha}$ diagram we also find statistical evidence of a non-linear flux-flux relationship. In particular, supported by the results of \citet{Houdebine1995} and \citet{Houdebine1997}, we argue that the \ion{Ca}{ii} H\&K emission flux increases monotonically with the stellar activity level, while the \halpha\ line initially goes into absorption and then is filled in by radiative emission processes. This suggests that at very low activity levels the \halpha\ absorption by filaments is more evident, while at higher activity levels the emission by plages dominates.

Searching for the physical origin of the scatter and trends in the $\rm F_{HK}$ vs. $\rm F_{H\alpha}$ diagram, we find that long-term activity cycles on year timescales play a minor role in the overall variability. Conversely, short-term activity explains $\gtrsim$60\% of the variance of the flux excess measurements.

After correcting the collected time series for year-long variability (whose analysis is not the scope of the present study), we find marginal statistical evidence that the slope of the $\rm F_{HK}$ vs. $\rm F_{H\alpha}$ relationship increases with decreasing stellar \teff. This is consistent with a scenario in which cooler stars tend to be less affected by chromospheric filaments as suggested by \citet{Meunier2009}, and as we found in \citet{Maldonado2016} with an independent analysis of the Balmer decrements of the same sample of stars.

We also find that the variance of the flux excess is an increasing function of stellar activity, and that even the quietest stars show some degree of variability. We attempt a rough time series analysis using the Pooled Variance approach suggested by \citet{Donahue1} and \citet{Donahue2}. We find evidence for rotation periods of the order of 10--40 days and active regions lifetime cycles longer than $\simeq$50 days. These findings are in agreement with previous results \citep{Donahue1,Donahue2,Reiners2012,Robertson2013} and are supported by a more detailed analysis we are currently preparing \citep{rotation}.

\begin{acknowledgements}

G. S.\ and I. P.\ acknowledge financial support from \lq\lq\ Accordo ASI--INAF\rq\rq\ n. 2013-016-R.0 Jul,9 2013.

J. M.\ acknowledges support from the Italian Ministry of Education, University, and Research through the PREMIALE WOW 2013 research project under grant \lq\lq Ricerca di pianeti intorno a stelle di piccola massa\rq\rq.

GAPS acknowledges support from INAF through the Progetti Premiali funding scheme of the Italian Ministry of Education, University, and Research.

J.I.GH. acknowledges financial support from the Spanish Ministry of Economy and Competitiveness (MINECO) under the 2013 Ram\'on y Cajal program MINECO RYC-2013-14875, and A.SM., J.I.GH., and R.R. also acknowledges financial support from the Spanish ministry project MINECO AYA2014-56359-P.

I.R.\ acknowledges support from the Spanish Ministry of Economy and Competitiveness (MINECO) through grant ESP2014-57495-C2-2-R.

This work is based on observations made with the Italian Telescopio Nazionale Galileo (TNG), operated on the island of La Palma by the Fundaci\'on Galielo Galilei of the Istituto Nazionale di Astrofisica (INAF) at the Spanish Observatorio del Roque de los Muchachos (ORM) of the Instituto de Astrof\'isica de Canarias (IAC).
\end{acknowledgements}


\begin{thebibliography}{}
\bibitem[Affer et al.(2016)]{Affer2016} Affer, L. 2016, \aap, accepted
\bibitem[Allard et al.(2011)]{Allard2011} Allard, F., Homeier, D., \& Freytag, B.\ 2011, 16th Cambridge Workshop on Cool Stars, Stellar Systems, and the Sun, 448, 91 
\bibitem[Anderson \& Darling(1952)]{Anderson1952} Anderson, T. W., Darling, D.~A. 1952. Annals of Mathematical Statistics 23: 193-212.
\bibitem[Baliunas et al.(1998)]{Baliunas1998} Baliunas, S.~L., Donahue, R.~A., Soon, W., \& Henry, G.~W.\ 1998, Cool Stars, Stellar Systems, and the Sun, 154, 153
\bibitem[Bochanski et al.(2007)]{Bochanski2007} Bochanski, J.~J., West, A.~A., Hawley, S.~L., \& Covey, K.~R.\ 2007, \aj, 133, 531 
\bibitem[Bonfils et al.(2013)]{Bonfils2013} Bonfils, X., Delfosse, X., Udry, S., et al.\ 2013, \aap, 549, A109 
\bibitem[Borsa et al.(2015)]{Borsa2015} Borsa, F., Scandariato, G., Rainer, M., et al.\ 2015, \aap, 578, A64
%\bibitem[Browning et al.(2010)]{Browning2010} Browning, M.~K., Basri, G., Marcy, G.~W., West, A.~A., \& Zhang, J.\ 2010, \aj, 139, 504 
\bibitem[Burnham \& Anderson(2002)]{Burnham2002} Burnham, K.~P., \& Anderson, D.~R. 2002. \textit{Model Selection and Multimodel Inference: A Practical Information-Theoretic Approach, 2nd ed.}, Springer-Verlag. ISBN 0-387-95364-7
%\bibitem[Caffau et al.(2011)]{Caffau2011} Caffau, E., Ludwig, H.-G., Steffen, M., Freytag, B., \& Bonifacio, P.\ 2011, \solphys, 268, 255
\bibitem[Cosentino et al.(2012)]{Cosentino2012} Cosentino, R., Lovis, C., Pepe, F., et al.\ 2012, \procspie, 8446, 84461V 
\bibitem[Covino et al.(2013)]{Covino2013} Covino, E., Esposito, M., Barbieri, M., et al.\ 2013, \aap, 554, A28 
\bibitem[Cram \& Mullan(1979)]{Cram1979} Cram, L.~E., \& Mullan, D.~J.\ 1979, \apj, 234, 579 
\bibitem[Cram \& Giampapa(1987)]{Cram1987} Cram, L.~E., \& Giampapa, M.~S.\ 1987, \apj, 323, 316 
\bibitem[Davenport et al.(2015)]{Davenport2015} Davenport, J.~R.~A., Hebb, L., \& Hawley, S.~L.\ 2015, \apj, 806, 212 
\bibitem[Donahue et al.(1997a)]{Donahue1} Donahue, R.~A., Dobson, A.~K., \& Baliunas, S.~L.\ 1997, \solphys, 171, 211 
\bibitem[Donahue et al.(1997b)]{Donahue2} Donahue, R.~A., Dobson, A.~K., \& Baliunas, S.~L.\ 1997, \solphys, 171, 191 
\bibitem[Frasca \& Catalano(1994)]{Frasca1994} Frasca, A., \& Catalano, S.\ 1994, \aap, 284, 883 
%\bibitem[Freytag et al.(2010)]{Freytag2010} Freytag, B., Allard, F., Ludwig, H.-G., Homeier, D., \& Steffen, M.\ 2010, \aap, 513, A19 
%\bibitem[Giampapa et al.(1982)]{Giampapa1982} Giampapa, M.~S., Worden, S.~P., \& Linsky, J.~L.\ 1982, \apj, 258, 740 
\bibitem[Giampapa et al.(1989)]{Giampapa1989} Giampapa, M.~S., Cram, L.~E., \& Wild, W.~J.\ 1989, \apj, 345, 536 
\bibitem[Gomes da Silva et al.(2011)]{Gomes2011} Gomes da Silva, J., Santos, N.~C., Bonfils, X., et al.\ 2011, \aap, 534, A30 
\bibitem[Gray(1992)]{Gray1992} Gray, D.~F.\ 1992, Camb.~Astrophys.~Ser., Vol.~20,,
%\bibitem[Henry et al.(1996)]{Henry1996} Henry, T.~J., Soderblom, D.~R., Donahue, R.~A., \& Baliunas, S.~L.\ 1996, \aj, 111,  
\bibitem[Henry et al.(2006)]{Henry2006} Henry, T.~J., Jao, W.-C., Subasavage, J.~P., et al.\ 2006, \aj, 132, 2360 
\bibitem[Herbig(1985)]{Herbig1985} Herbig, G.~H.\ 1985, \apj, 289, 269 
\bibitem[Houdebine et al.(1995)]{Houdebine1995} Houdebine, E.~R., Doyle, J.~G., \& Koscielecki, M.\ 1995, \aap, 294, 773 
\bibitem[Houdebine \& Stempels(1997)]{Houdebine1997} Houdebine, E.~R., \& Stempels, H.~C.\ 1997, \aap, 326, 1143 
\bibitem[Houdebine(2011)]{Houdebine2011} Houdebine, E.~R.\ 2011, \mnras, 411, 2259 
%\bibitem[Isobe et al.(1990)]{Isobe1990} Isobe, T., Feigelson, E.~D., Akritas, M.~G., \& Babu, G.~J.\ 1990, \apj, 364, 104 
\bibitem[Kendall(1938)]{Kendall1938} Kendall, M. (1938), Biometrika, 30 (1–2), 81.
\bibitem[Lanza et al.(2004)]{Lanza2004} Lanza, A.~F., Rodon{\`o}, M., \& Pagano, I.\ 2004, \aap, 425, 707 
\bibitem[Legendre \& Legendre(1983)]{Legendre1998} Legendre, P., Legendre, L., 1998, Numerical Ecology, Elsevier
\bibitem[Lovis \& Pepe(2007)]{Lovis2007} Lovis, C., \& Pepe, F.\ 2007, \aap, 468, 1115 
\bibitem[Maldonado et al.(2010)]{Maldonado2010} Maldonado, J., Mart{\'{\i}}nez-Arn{\'a}iz, R.~M., Eiroa, C., Montes, D., \& Montesinos, B.\ 2010, \aap, 521, A12 
\bibitem[Maldonado et al.(2015)]{Maldonado2015} Maldonado, J., Affer, L., Micela, G., et al.\ 2015, \aap, 577, A132 
\bibitem[Maldonado et al.(2016)]{Maldonado2016} Maldonado, J. et al.\ 2016, \aap, submitted
%\bibitem[Mart{\'{\i}}nez-Arn{\'a}iz et al.(2010)]{Martinez2010} Mart{\'{\i}}nez-Arn{\'a}iz, R., Maldonado, J., Montes, D., Eiroa, C., \& Montesinos, B.\ 2010, \aap, 520, A79 
\bibitem[Mart{\'{\i}}nez-Arn{\'a}iz et al.(2011)]{Martinez2011} Mart{\'{\i}}nez-Arn{\'a}iz, R., L{\'o}pez-Santiago, J., Crespo-Chac{\'o}n, I., \& Montes, D.\ 2011, \mnras, 417, 3100 
\bibitem[Meunier \& Delfosse(2009)]{Meunier2009} Meunier, N., \& Delfosse, X.\ 2009, \aap, 501, 1103 
\bibitem[Montes et al.(1995)]{Montes1995} Montes, D., Fernandez-Figueroa, M.~J., de Castro, E., \& Cornide, M.\ 1995, \aap, 294,  
\bibitem[Newton et al.(2016)]{Newton2016} Newton, E.~R., Irwin, J., Charbonneau, D., et al.\ 2016, \apj, 821, 93 
%\bibitem[Noyes et al.(1984)]{Noyes1984} Noyes, R.~W., Hartmann, L.~W., Baliunas, S.~L., Duncan, D.~K., \& Vaughan, A.~H.\ 1984, \apj, 279, 763
%\bibitem[Pagano(2013)]{Pagano2013} Pagano, I.\ 2013, Planets, Stars and Stellar Systems.~Volume 4: Stellar Structure and Evolution, 485 
\bibitem[Pasquini \& Pallavicini(1991)]{Pasquini1991} Pasquini, L., \& Pallavicini, R.\ 1991, \aap, 251, 199 
\bibitem[Perger et al.(2016)]{Perger2016} Perger, M.\ 2016, Review, XX, YY
\bibitem[Rauscher \& Marcy(2006)]{Rauscher2006} Rauscher, E., \& Marcy, G.~W.\ 2006, \pasp, 118, 617 
\bibitem[Reid et al.(2002)]{Reid2002} Reid, I.~N., Gizis, J.~E., \& Hawley, S.~L.\ 2002, \aj, 124, 2721 
\bibitem[Reiners et al.(2012)]{Reiners2012} Reiners, A., Joshi, N., \& Goldman, B.\ 2012, \aj, 143, 93 
\bibitem[Robertson et al.(2013)]{Robertson2013} Robertson, P., Endl, M., Cochran, W.~D., \& Dodson-Robinson, S.~E.\ 2013, \apj, 764, 3 
\bibitem[Robertson et al.(2015)]{Robertson2015} Robertson, P., Endl, M., Henry, G.~W., et al.\ 2015, \apj, 801, 79 
\bibitem[Robinson et al.(1990)]{Robinson1990} Robinson, R.~D., Cram, L.~E., \& Giampapa, M.~S.\ 1990, \apjs, 74, 891 
\bibitem[Rutten et al.(1989)]{Rutten1989} Rutten, R.~G.~M., Zwaan, C., Schrijver, C.~J., Duncan, D.~K., \& Mewe, R.\ 1989, \aap, 219, 239 
\bibitem[Stauffer \& Hartmann(1986)]{Stauffer1986} Stauffer, J.~R., \& Hartmann, L.~W.\ 1986, Cool Stars, Stellar Systems and the Sun, 254, 58 
\bibitem[Stelzer et al.(2012)]{Stelzer2012} Stelzer, B., Alcal{\'a}, J., Biazzo, K., et al.\ 2012, \aap, 537, A94 
\bibitem[Stelzer et al.(2013)]{Stelzer2013} Stelzer, B., Frasca, A., Alcal{\'a}, J.~M., et al.\ 2013, \aap, 558, A141 
\bibitem[Strassmeier et al.(1990)]{Strassmeier1990} Strassmeier, K.~G., Fekel, F.~C., Bopp, B.~W., Dempsey, R.~C., \& Henry, G.~W.\ 1990, \apjs, 72, 191
\bibitem[Su{\'a}rez Mascare{\~n}o et al.(2015)]{Suarez2015} Su{\'a}rez Mascare{\~n}o, A., Rebolo, R., Gonz{\'a}lez Hern{\'a}ndez, J.~I., \& Esposito, M.\ 2015, \mnras, 452, 2745 
\bibitem[Su{\'a}rez Mascare{\~n}o et al.(2016, in prep)]{rotation} Su{\'a}rez Mascare{\~n}o, A., in preparation
\bibitem[Thatcher \& Robinson(1993)]{Thatcher1993} Thatcher, J.~D., \& Robinson, R.~D.\ 1993, \mnras, 262, 1 
\bibitem[Vidotto et al.(2013)]{Vidotto2013} Vidotto, A.~A., Jardine, M., Morin, J., et al.\ 2013, \aap, 557, A67 
\bibitem[Walkowicz \& Hawley(2009)]{Walkowicz2009} Walkowicz, L.~M., \& Hawley, S.~L.\ 2009, \aj, 137, 3297 
\bibitem[West et al.(2004)]{West2004} West, A.~A., Hawley, S.~L., Walkowicz, L.~M., et al.\ 2004, \aj, 128, 426 
\bibitem[West et al.(2011)]{West2011} West, A.~A., Morgan, D.~P., Bochanski, J.~J., et al.\ 2011, \aj, 141, 97 
%\bibitem[Zechmeister \& Kurster(2009)]{scargle} Zechmeister, M., \& K\"urster, M. 2009, \aap, 496, 577





\end{thebibliography}
\end{document}